\documentclass[12pt,thmsa]{article}
\usepackage{amssymb}

\usepackage{sw20lart}

\input{tcilatex}
\begin{document}

\author{Emilio Santos \\
%EndAName
Departamento de F\'{i}sica.\\
University of Cantabria. Santander. Spain\\
email: santose@unican.es}
\title{The quantum theory of the electromagnetic field in the Weyl-Wigner
representation as a local realistic model}
\date{Jan,31,2024}
\maketitle
\tableofcontents

\begin{abstract}
I revisit the Wigner (or Weyl-Wigner, WW) representation of the quantum
electromagnetic field. I show that, assuming that Fock states are just
mathematical concepts devoid of physical reality, WW suggests a realistic
interpretation which turns out to be (classical) Maxwell theory with the
assumption that there is a random radiation filling space, the vacuum field.
I elucidate why, in sharp constrast, non-relativistic quantum mechanics of
particles does not admit a realistic interpretation via WW. I interpret
experiments involving entangled light beams within WW, in particular optical
tests of Bell inequalities. I show that WW provides clues in order to
construct local model for those experiments. I give arguments why Bell
defintion of local realism is not general enough.
\end{abstract}

\section{Introduction}

The loophole-free violation of Bell inequalities \cite{Bell64}, \cite{Bell}
in experiments with entangled photon pairs \cite{Shalm}, \cite{Giustina}, 
\cite{BIG}, has been claimed the death by experiments for local realism \cite
{Wiseman}, \cite{Aspect}. However local realistic models for those
experiments have been published \cite{Frontiers}, \cite{FOOP}, \cite{book},
whose predictions agree with quantum theory for low detection efficiency,
although they are unable to violate a Bell inequality. Indeed the models
cannot be extended trivially until the efficiencies needed for the
violation, that is more than 67\% for the quoted experiments \cite{Shalm}, 
\cite{Giustina}. Those models afford indications about the possible
existence of new loopholes not yet discovered, although the subject has been
carefully studied and there is a consensus that no loophole remains. Anyway
the conflict between the violation of a Bell inequality and relativistic
causality is awful so that new research is worthwhile.

In this paper I will study the subject in more detail via the quantum theory
of the electromagnetic field in the Wigner representation (or Weyl-Wigner,
WW). It suggets explicit local models for optical experiments involving
entangled photon pairs. I will discuss to what extent the existence of these
models might provide a better understanding of the conflict between
experiments and relativistic causality.

In the next section 2 I briefly revisit the well known quantum theory of the
electromagnetic (EM) field, firstly in Hilbert space, then in the Wigner
representation, stressing its most characteristic traits. Section 3 deals
with correlation experiments involving entangled photon pairs, analized
firstly within the standard Hilbert space formalism and then in the WW
representation. In section 4 I will show that the Wigner representation of
the EM quantum field provides an actual realistic local model for the said
correlation experiments. In section 5 I shall study whether the said local
model fits, or does not fit, in Bell\'{}s celebrated formulation of local
realism \cite{Bell}.

\section{Quantum theory of the electromagnetic field}

\subsection{Hilbert space formulation}

The standard method to describe a field is to expand it in plane waves or,
more generally, normal modes. In the classical theory of the EM field the
amplitudes of the modes are conveniently written using two complex
conjugated quantities (c-numbers) $\left\{ a_{j},a_{j}^{*}\right\} ,$ where $%
j$ labels a mode. Here I will study that field in the Coulomb gauge because
I believe that, in order to get an intuitive picture, it is more appropriate
than the Lorentz gauge, which is more common and useful in the formal study
of relativistic quantum fields. In the Coulomb gauge the quantity expanded
in normal modes is the vector potential \textbf{A}(\textbf{x},t). It should
be supplemented with a scalar potential taking account of the instantaneous
electrostatic interaction between charges, but the effect of the latter is
trivial and it will be ignored in the following.

In free space the expansion may be written 
\begin{equation}
\mathbf{A}(\mathbf{x},t)=\sum_{l}a_{l}\mathbf{\varepsilon }_{l}\exp \left( i%
\mathbf{k}_{l}\mathbf{\cdot x-}i\omega _{l}t\right) +\func{complex}conjugate,
\label{A}
\end{equation}
where $\mathbf{k}_{l}$ is the wavevector and $\mathbf{\varepsilon }_{l}$ the
polarization vector of a mode with frequency $\omega _{l}=c\left| \mathbf{k}%
_{l}\right| $. Standard quantization of the field consists of introducing a
Hilbert space and promoting the classical amplitudes to be operators $%
\left\{ \hat{a}_{j},\hat{a}_{j}^{\dagger }\right\} $\ acting on that space.
They fulfil the commutation rules 
\begin{equation}
\left[ \hat{a}_{j},\hat{a}_{k}\right] =\left[ \hat{a}_{j}^{\dagger },\hat{a}%
_{k}^{^{\dagger }}\right] =0,\left[ \hat{a}_{j},\hat{a}_{k}^{\dagger
}\right] =\delta _{jk},  \label{com}
\end{equation}
$\delta _{jk}$\ being Kronecker delta. The vacuum state is represented by a
vector $\mid 0\rangle $\ defined by 
\begin{equation}
\hat{a}_{j}\mid 0\rangle =0\Rightarrow \langle 0\mid \hat{a}_{j}^{\dagger
}=0,\text{ for all radiation modes,}  \label{0}
\end{equation}
$0$\ being here the null vector in the HS.

Then it is assumed that \textbf{pure states} of the field correspond to
vectors which may be got by application of the ``creation'' operators of
photons $\left\{ \hat{a}_{j}^{\dagger }\right\} $ to the vacuum state. For
instance $\hat{a}_{j}\dagger \mid 0\rangle $ is the vector of HS
representing a single photon of kind $j$, the vector $\hat{a}_{j}\dagger 
\hat{a}_{k}\dagger \mid 0\rangle $ represents a two-photon state, etc. The
set of all these states is named Fock space. It is currently admitted that
any linear combination of state-vectors belonging to Fock space represents
also a possible pure state. Hence all pure states, $\left\{ \mid \psi
\rangle \right\} ,$ could be written as functions of the creation operators
acting on the vacuum state, that is 
\begin{equation}
\mid \psi \rangle =\hat{f}\mid 0\rangle ,\hat{f}\equiv
c_{0}+\sum_{j}\sum_{n}c_{jn}\hat{a}_{j}^{\dagger n},  \label{stateHS}
\end{equation}
where $j$ labels a radiation mode and $n$ are natural numbers, $c_{0}$ and $%
c_{jn}$ being complex numbers, with the constraint that the vector $\mid
\psi \rangle $ is normalized. \textbf{Mixed states}, which might be
represented by density operators, correspond to probability distributions of
pure states.

\textbf{Observables} are self-adjoint operators in HS. They are usually
defined as functions of the elementary operators $\left\{ \hat{a}_{j},\hat{a}%
_{j}^{\dagger }\right\} .$ In particular the free field Hamiltonian is 
\begin{equation}
\hat{H}_{HS}=\frac{1}{2}%TCIMACRO{\UNICODE[m]{0x127}}
%BeginExpansion
\rlap{\protect\rule[1.1ex]{.325em}{.1ex}}h%
%EndExpansion
\sum_{j}\omega _{j}(\hat{a}_{j}^{\dagger }\hat{a}_{j}+\hat{a}_{j}\hat{a}%
_{j}^{\dagger })=%TCIMACRO{\UNICODE[m]{0x127}}
%BeginExpansion
\rlap{\protect\rule[1.1ex]{.325em}{.1ex}}h%
%EndExpansion
\sum_{j}\omega _{j}(\hat{a}_{j}^{\dagger }\hat{a}_{j}+\frac{1}{2}).
\label{observable}
\end{equation}
However with this definition the energy of the vacuum state $\mid 0\rangle $
is not zero (in fact it diverges) whence it is standard to redefine the
Hamiltonian with a different order for the operators. Using that `normal
ordering rule' the Hamiltonian becomes 
\begin{equation}
\hat{H}_{HS}^{normal}=%TCIMACRO{\UNICODE[m]{0x127}}
%BeginExpansion
\rlap{\protect\rule[1.1ex]{.325em}{.1ex}}h%
%EndExpansion
\sum_{j}\omega _{j}\hat{a}_{j}^{\dagger }\hat{a}_{j},  \label{normalorder}
\end{equation}
which gives zero energy for the vacuum state$.$

The \textbf{evolution} with time may be got from the Hamiltonian via the
Heisenberg equation.

Up to here everything is well known and looks uncontroversial. However the
meaning of states and observables in quantum theory is confused and it
requires a careful studied which is made in the next subsection 2.2.

\subsection{A digression on the interpretation of quantum theory}

The concepts introduced in the quantum formalism (states, pure or mixed,
observables, Hamiltonian) are well defined \textit{mathematical} objects
within the Hilbert space theory, and the evolution is a continuous mapping
of that space. However the formalism alone does not allow interpreting the
observations or experiments. It is necessary to introduce a connection
between the formalism and the empirical facts which is named \textbf{%
measurement theory}. This is a novelty in comparison with classical
mechanics of particles, where no measurement theory is needed or it is
trivial. In fact in classical mechanics all measurements may be reduced to
position measurements including bulk matter that may be treated as a
continuous set of (infinitely many) particles. For some other branches of
classical physics, like thermodynamics, a more sophisticated connection is
required, but this is made by \textit{particular protocols}, not via a 
\textit{general} measurement theory. For instance a protocol in order to
prepare a sample of gas at a given temperature and pressure, or for the
measurement of air temperature. Similarly, protocols may exist for other
particular measurements or observations. The existence of protocols for
preparation and measurements is also common in chemistry and other branches
of science. But in classical physics (or other sciences of nature) no
general statements exists like ``the measurement of an observable (operator)
\^{M} gives rise to one of its eigenvalues, say $m,$ where 
\[
\hat{M}\mid \psi _{m}\rangle =m\mid \psi _{m}\rangle , 
\]
$\mid \psi _{m}\rangle $ being one of the eigenvectors of $\hat{M},$ here
assumed non-degenerate for simplicity.

The mere existence of a \textit{general} measurement theory has been
criticized by philosophers of science like Karl Popper \cite{Popper} or
Mario Bunge \cite{Bunge}. Also by Albert Einstein who supported that
``science should be concerned with what nature does, not with the relation
between nature and human observers'' \cite{Einstein}. In my view the
problems for the interpretation of quantum theory \cite{Handbook} come from
the need, and the existence, of a general measurement theory. Indeed the
formalism plus the measurement theory induce people to ascribe \textit{%
physical} reality to all \textit{mathematical} concepts, e.g. single-photon
states in the quantum theory of the electromagnetic field. Then problems
arise like the dual nature of light, which is assumed to \textit{consist} of
particles and fields (waves) simultaneously, two contradictory images
indeed. We might rather say that, in different contexts, light \textit{%
behaves} in a way that \textit{suggests} either waves or particles.

I believe that the quantum formalism together with the measurement theory is
an algoritm (extremely efficient!) for the calculation of the results of
experiments. It should inform us about what nature does (see e.g. the
initial paragraph of the celebrated EPR article \cite{EPR} ) but it does
not. The present formalism plus the measurement theory leads to an
unavoidable pragmatic approach, which was actually the germ of the
Copenhagen interpretation. Indeed for some people Copenhagen's is actually
the statement that no interpretation is required, we need just working rules
leading to good agreement with experiments.

After this general digression I may discuss briefly two illustrative
experiments that have been considered to support the particle behaviour of
light: by Grangier, Roger and Aspect \cite{Grangier}, and by Lvovsky et al. 
\cite{Lvovsky}. In the former a weak light beam consisting of single-photon
states is sent to a beam-splitter BS1. The single photon state is claimed to
be guaranteed by the fact that the measurements are made in coincidence with
one detection event in the partner beam produced in \textit{parametric
down-conversion} (see subsection 3.1 for the concept). The result is that no
coincidence counts are detected in the two beams emerging from BS1, which is
claimed to show that photons cannot be divided, that is every photon goes to
one of the outgoing channels of BS1. However when these fields are
recombined by appropriate mirrors and a new beam splitter BS2, the incident
beam may be reconstructed, as is shown by an interference effectin a beam
emerging from BS2. Indeed there is either constructive or destructive
interference depending on the relative length of the two paths from BS1 to\
BS2, which is shown by the change of detection rate in a detector placed in
front of one of outgoing channel of BS2. The recombination is claimed to
show the wave character of light. It is the case that the results of the
experiment were explained, almost 40 years ago \cite{MS}, with a very simple
realistic (classical-like) model involving the vacuum radiation. That model
has been reproduced in several reviews \cite{Handbook}, \cite{Foundations}.
Nevertheless the experiment is still quoted in popularizations of quantum
theory as a paradigmatic example of the impossibility of an intuitive
picture of the (quantum) microworld, see e.g. \cite{Rovelli}.

In another celebrated experiment Lvovsky et al. \cite{Lvovsky} were able to
reconstruct experimentally the (non-positive) Wigner function of a
single-photon Fock state, using the method of phase-randomized pulsed
optical homodyne tomography. The single photon state is claimed to be
guaranteed again by the measurement being made in coincidence with the
partner beam, both beams produced via parametric down-conversion. The
results show classically impossible negative values around the origin of the
phase space. The experiment is a beautiful confirmation of quantum theory,
but it does not prove the impossibity of an interpretation via some
realistic (classical-like) theory. Indeed the \textit{actually measured
results} are rates of counts in a homodyne detector conditional to counts in
another (trigger) detector, and these rates are what should be reproduced by
a realistic model, without the need of any reference to photons. Exhibiting
a detailed realistic model of the experiment is out of the scope of the
present paper, which deals with local models for correlation experiments.

In summary a realistic theory should predict \textit{the results} of the
experiment, but not necessarily confirm the\textit{\ }(particle or wave)
nature of light \textit{suggested} by these results. In general it is an
epistemological principle that agreement of an experiment with a theory does
not imply that there is no other theory in agreement too. Furthermore a
single success, or many successes, of a theory do not completely confirm the
theory, although a single disagreement does refute it \cite{Popper}.

I conclude that the Hilbert space formalism plus the measurement theory make
quantum mechanics a quite efficient algoritm for the prediction of the
results of experiments. However we should refrain from searching for a
satisfactory picture of reality resting on the standard (Hilbert space)
formulation. Indeed the interpretational problems of quantum theory are well
known and long standing (see e.g. \cite{Handbook}). A solution is to use an
alternative formulation able to provide a picture of reality. I propose that
the Wigner representation is a good candidate, at least for the quantized
electromagnetic field. It is physically equivalent to the common HS
formalism in the sense that it makes the same predictions for experiments,
with some cautions as commented on below.

\subsection{The Weyl-Wigner formalism in quantum mechanics and field theory}

The formalism derives from the transform that H. Weyl introduced in 1928 
\cite{Weyl} for non-relativistic quantum mechanics of particles (QM), and
the work of Wigner in 1932 \cite{Wigner}. It is usually known as ``Wigner
representation'' \cite{Scully}, \cite{Zachos} (which for the electromagnetic
field I have labeled WW, which will be used for short in the following). In
the Wigner representation of QM both the states of systems (named Wigner
functions of the states) and the observables appear as functions in phase
space (i.e. the space of positions and momenta of the particles). WW is a
valid alternative to the standard Hilbert space (HS) for calculations, but
it does not offer a picture of reality. In fact the said phase-space
functions are positive but for a small fraction of the quantum states i.e.
those whose Wigner function is Gaussian \cite{Soto}. For this reason it is
widely rejected as a possible formulation able to provide a realistic
interpretation for QM. In sharpt contrast I propose that WW is valid for
fields, at least for the quantum electromagnetic field.

In my opinion the reason why the Wigner representation does not allow a
realistic interpretation of the QM for particles is that quantum particles
are not simple, but quite complex, objects. Indeed a quantum particle
usually consists of a \textit{bare} particle \textit{dressed }by many
interacting fields with the consequence that the physical mass and charge of
the particle are quite different from the bare quantities. Therefore the
state of the particle cannot be described by just position and momentum
variables, but it depends in a complex manner from the dressing fields. When
there are several particles at a small distance the dressing fields of all
particles are involved and they interact, whence the phase-space
representation losses any meaning.

As is well known the Wigner representation may be easily extended to the
quantized electromagnetic field. At a difference with QM the Wigner
representation does suggest a quite clear realistic picture for the field.
In the following I recall briefly that formalism, which has been studied
elsewhere \cite{Frontiers}, \cite{FOOP}, \cite{EPJP}, \cite{book}. As
commented above WW is currently rejected as a possible realistic
interpretation of QM, but I believe that in sharp contrast the WW
representation for the EM field is fair by the following reasons:

1. In quantized EM field (QEM) the ground has a Wigner function which is
positive definite, see eq.$\left( \ref{1}\right) $ below. In contrast in QM
the ground state of a system of particles has frequently a Wigner function
which is not positive.

2. The evolution of the QEM field either free or interacting with
macroscopic bodies (as in optical experiments) is governed by the classical
Maxwell-Lorentz electrodynamics, which preserve positivity of the
probability distributions. This is not the case in QM$.$

3. The states of QEM are most times produced in practice by the action of
macroscopic devices on the vacuum, which would give rise to states with
positive Wigner function. This happens for instance in spontaneous
parametric down conversion leading to ``entangled photon pairs'', to be
studied in section 4.1.

4. In the Hilbert space interpretation of experiments, n-photon states
usually appear but at intermediate stages of the calculation, whose
transform via eq.$\left( \ref{Wtransform}\right) $ might not be positive$.$
However I argue that there is no need to ascribe physical reality to these
intermediate mathematical elements in the calculation.

\subsection{Weyl-Wigner representation of the EM field}

The Weyl (or Weyl-Wigner) transform, which is invertible, allows the passage
from WW to HS and the inverse transform from HS to WW. The latter is more
relevant for us, and it may be written

\begin{eqnarray}
W_{\hat{M}} &=&T_{W}\left[ \hat{M}\right] \equiv \frac{1}{(2\pi ^{2})^{n}}%
\prod_{j=1}^{n}\int_{-\infty }^{\infty }d\lambda _{j}\int_{-\infty }^{\infty
}d\mu _{j}\exp \left[ -2i\lambda _{j}\mathrm{Re}a_{j}-2i\mu _{j}\mathrm{Im}%
a_{j}\right]  \nonumber \\
&&\times Tr\left\{ \hat{M}\exp \left[ i\lambda _{j}\left( \hat{a}_{j}+\hat{a}%
_{j}^{\dagger }\right) +\mu _{j}\left( \hat{a}_{j}-\hat{a}_{j}^{\dagger
}\right) \right] \right\} ,  \label{Wtransform}
\end{eqnarray}
where $T_{W}\left[ \hat{M}\right] $ stands for (inverse) Weyl transform of
the HS operator $\hat{M}.$ (For the sake of clarity I write operators in a
Hilbert space with a `hat', e. g. $\hat{a}_{j},\hat{a}_{j}^{\dagger },$ and
numerical (c-number) amplitudes without `hat', e. g. $a_{j},a_{j}^{*}).$ The
result obtained, $W_{\hat{M}}\left( \left\{ a_{j},a_{j}^{*}\right\} \right)
, $ is a function of (c-number) amplitudes. The most relevant properties of
WW, required for section 3, are well known and may be seen in the
references, e.g. \cite{FOOP}, whence I will skip details.

\textbf{Transform of products of field amplitudes.} Eq.$\left( \ref
{Wtransform}\right) $ allows deriving the product of amplitudes in WW for
any HS product of creation or annihilation operators. If we have a product
of WW amplitudes like $a_{j}^{m}a_{j}^{*n}$ the HS counterpart is 
\begin{equation}
a_{j}^{m}a_{j}^{*n}\rightarrow \left( \hat{a}_{j}^{m}\hat{a}_{j}^{\dagger
n}\right) _{sym},  \label{sym}
\end{equation}
where \textit{sym} stands for symmetric and it means writing a sum of the $%
m+n$ operators in all possible orderings and then dividing by the number of
terms, which is $\left( m+n\right) !/(m!n!)$.

\textbf{The ``vacuum'' state. }The vacuum state\textbf{\ }is represented in
HS either by the state vector $\mid 0\rangle ,$ defined by eq.$\left( \ref{0}%
\right) ,$ or by the density operator 
\begin{equation}
\hat{\rho}=\mid 0\rangle \langle 0\mid .  \label{2d}
\end{equation}
If this density operator is inserted in place of $\hat{M}$ in eq.$\left( \ref
{Wtransform}\right) $ we get after some algebra the WW expression for the
vacuum state, that is

\begin{equation}
W_{0}=\prod_{j}\frac{2}{\pi }\exp \left( -2\left| a_{j}\right| ^{2}\right) ,
\label{1}
\end{equation}
which I write normalized for the integration with respect to $\prod_{j}d%
\func{Re}a_{j}d\func{Im}a_{j}.$ Taking the last eq.$\left( \ref{Hamiltonians}%
\right) $ (see below) into account eq.$\left( \ref{1}\right) $ easily leads
to the following distribution of the energy amongst the radiation modes 
\begin{equation}
W_{E}=\prod_{j}\frac{2}{%TCIMACRO{\UNICODE[m]{0x127}}
%BeginExpansion
\rlap{\protect\rule[1.1ex]{.325em}{.1ex}}h%
%EndExpansion
\omega _{j}}\exp \left( -\frac{2E_{j}}{%TCIMACRO{\UNICODE[m]{0x127}}
%BeginExpansion
\rlap{\protect\rule[1.1ex]{.325em}{.1ex}}h%
%EndExpansion
\omega _{j}}\right) ,  \label{W0}
\end{equation}
where $E_{j}$ is the energy of mode $j$ and the normalization is appropriate
for integration with respect to $\prod_{j}dE_{j}$.

Hence a plausible interpretation of the vacuum state $W_{0}$ of the field in
the WW formalism i\textit{s a real random radiation, that is a stochastic
field, filling space with the distribution eq.}$\left( \ref{1}\right) $ with
a mean energy $\frac{1}{2}%TCIMACRO{\UNICODE[m]{0x127}}
%BeginExpansion
\rlap{\protect\rule[1.1ex]{.325em}{.1ex}}h%
%EndExpansion
\omega _{j}$ per normal mode. A zero-point energy for material harmonic
oscillators was proposed as early as 1912 in the second Planck's radiation
law, and extended to radiation by W. Nernst in 1916. It is currently named
zeropoint energy or field (ZPF), and treated as a formal ingredient. Anyway
the existence of vacuum fluctuations is widely accepted and it has a
fundamental role in some calculations, e.g. renormalization technques of
relativistic QED. The radiation eq.$\left( \ref{1}\right) $ corresponds to
the vacuum state in HS and therefore it is the ground state of the field in
WW. Then in WW the vacuum is not empty but filled with a random radiation.

\textit{Comment on units. }Comparison of eqs.$\left( \ref{1}\right) $ and $%
\left( \ref{W0}\right) $ shows that we may identify 
\begin{equation}
\left| a_{j}\right| ^{2}=E_{j}/%TCIMACRO{\UNICODE[m]{0x127}}
%BeginExpansion
\rlap{\protect\rule[1.1ex]{.325em}{.1ex}}h%
%EndExpansion
\omega _{j},  \label{intensity}
\end{equation}
that is the modulus square of the amplitude $a_{j}$ may be considered an
energy measured in units $%TCIMACRO{\UNICODE[m]{0x127}}
%BeginExpansion
\rlap{\protect\rule[1.1ex]{.325em}{.1ex}}h%
%EndExpansion
\omega _{j},$ or in (dimensionless) number of photons.

\textbf{Other states. }If we attempted to parallel HS as closely as possible
we might define as states in WW the Weyl transform $\left( \ref{Wtransform}%
\right) $ of the set of states defined in HS, that is $\left( \ref{stateHS}%
\right) .$ However our interest in the WW formalism is to provide a
realistic interpretation for the quantized EM field, and most states in $%
\left( \ref{stateHS}\right) $ would give rise to Wigner\ states having a
non-positive distribution, which could not be interpreted as a probability
density. A typical example is the one-photon state. Therefore I propose that
all physical states in WW should correspond to (positive definite)
probability distributions for the field. The question is whether that set of
states are enough for the interpretation of \textit{all possible}
experiments. I believe it is, but a complete proof is obviously not possible
because the set of all possible experiments is not well defined.

I do not support the claim, frequently made, that the WW representation of
the electromagnetic \textit{quantized} field cannot admit a realistic
interpretation because there are states which have been shown \textit{%
empirically} to possess a non-positive Wigner function. Examples are the
experiments performed by Grangier et al.\cite{Grangier} and by Lvovsky et
al. \cite{Lvovsky} commented on section 2.2 with the conclusion that they do
not prevent a realistic interpretation.

\textbf{Observables. }Eq.$\left( \ref{Wtransform}\right) $ allows getting
the WW counterparts of the observables in the HS formalism. In particular
the free field Hamiltonians in the HS and WW formalisms are, respectively, 
\begin{equation}
\hat{H}_{HS}=%TCIMACRO{\UNICODE[m]{0x127}}
%BeginExpansion
\rlap{\protect\rule[1.1ex]{.325em}{.1ex}}h%
%EndExpansion
\sum_{j}\omega _{j}(\hat{a}_{j}^{\dagger }\hat{a}_{j}+\frac{1}{2})=\frac{1}{2%
}%TCIMACRO{\UNICODE[m]{0x127}}
%BeginExpansion
\rlap{\protect\rule[1.1ex]{.325em}{.1ex}}h%
%EndExpansion
\sum_{j}\omega _{j}(\hat{a}_{j}^{\dagger }\hat{a}_{j}+\hat{a}_{j}\hat{a}%
_{j}^{\dagger }),H_{WW}=%TCIMACRO{\UNICODE[m]{0x127}}
%BeginExpansion
\rlap{\protect\rule[1.1ex]{.325em}{.1ex}}h%
%EndExpansion
\sum_{j}\omega _{j}\left| a_{j}\right| ^{2}.  \label{Hamiltonians}
\end{equation}
However in the HS formalism it is common to change the order of the
operators putting the annihilation to the right, see eq.$\left( \ref
{normalorder}\right) $. Using this `normal ordering rule' the HS and WW
Hamiltonians become, respectivly, 
\begin{equation}
\hat{H}_{HS}^{normal}=%TCIMACRO{\UNICODE[m]{0x127}}
%BeginExpansion
\rlap{\protect\rule[1.1ex]{.325em}{.1ex}}h%
%EndExpansion
\sum_{j}\omega _{j}\hat{a}_{j}^{\dagger }\hat{a}_{j},H_{WW}^{normal}= 
%TCIMACRO{\UNICODE[m]{0x127}}
%BeginExpansion
\rlap{\protect\rule[1.1ex]{.325em}{.1ex}}h%
%EndExpansion
\sum_{j}\omega _{j}\left( \left| a_{j}\right| ^{2}-\frac{1}{2}\right) ,
\label{Hnormal}
\end{equation}
where I have taken into account eq.$\left( \ref{sym}\right) .$

\textbf{Evolution. }It is remarkable that the evolution\textbf{\ }of the 
\textit{quantized} free EM field within the WW formalism is just \textit{the
classical (Maxwell) evolution}. In fact the evolution of a field in the WW
formalism is given by the Moyal equation 
\begin{eqnarray}
\frac{\partial W_{\hat{M}}}{\partial t} &=&2\{\sin \left[ \frac{1}{4}\left( 
\frac{\partial }{\partial \mathrm{Re}a_{j}^{\prime }}\frac{\partial }{%
\partial \mathrm{Im}a_{j}^{\prime \prime }}-\frac{\partial }{\partial 
\mathrm{Im}a_{j}^{\prime }}\frac{\partial }{\partial \mathrm{Re}%
a_{j}^{\prime \prime }}\right) \right] \smallskip  \nonumber \\
&&\times W_{\hat{M}}\left\{ a_{j}^{\prime },a_{j}^{*\prime },t\right\}
H\left( a_{j}^{\prime \prime },a_{j}^{*\prime \prime }\right) \}_{a_{j}},
\label{Moyal}
\end{eqnarray}
where $\left\{ {}\right\} _{a_{j}}$ means making $a_{j}^{\prime
}=a_{j}^{\prime \prime }=a_{j}$ and $a_{j}^{*\prime }=a_{j}^{*\prime \prime
}=a_{j}^{*}$ after performing the derivatives. For the free electromagnetic
field the Hamiltonian $H\left( a_{j},a_{j}^{*}\right) $ is quadratic in the
amplitudes whence $\sin x$ reduces to $x$ in eq.$\left( \ref{Moyal}\right) ,$
$x$ being the square bracket. This means that the Moyal bracket becomes the
Poisson bracket of classical evolution. For details see e.g. Ref. \cite{EPJP}
.

\textbf{Expectation value }of the observable $\hat{O}$ in the state $\hat{%
\rho}$ reads $Tr(\hat{\rho}\hat{O})$, or in particular $\langle \psi \mid 
\hat{O}\mid \psi \rangle ,$ in the HS formalism. The counterpart in the WW
formalism leads to the integral of the product of two functions of the
amplitudes, that is 
\begin{equation}
Tr(\hat{\rho}\hat{O})=\int W_{\hat{\rho}}\left\{ a_{j},a_{j}^{*}\right\} W_{%
\hat{O}}\left\{ a_{j},a_{j}^{*}\right\} \prod_{j}d\mathrm{Re}a_{j}d\mathrm{Im%
}a_{j},  \label{expect}
\end{equation}
where $W_{\hat{\rho}}$ and $W_{\hat{O}}$ are the counterparts of a density
operator, $\hat{\rho},$ and a quantum observable $\hat{O}.$ A particular
case of eq.$\left( \ref{expect}\right) $ is the \textit{vacuum} expectation
value where $W_{\hat{\rho}}$ becomes $W_{0}.$ Eq.$\left( \ref{expect}\right) 
$ proves that the predictions of both HS and WW formalisms are identical for
the same states and observables, that is both formalisms represent the same 
\textit{physical theory} for the quantized EM field, although mathematically
the formalisms are quite different.

\textbf{Energy vacuum expectation. }As an illustration let us calculate the
expectation value of the energy within WW for a general state $W_{\phi }$ .
In order to agree with HS predictions we shall use the normally ordered
Hamiltonian eq.$\left( \ref{Hnormal}\right) .$ For a state $\mid \phi
\rangle $ of the field the calculation, firstly in HS then in WW, may be
written 
\begin{eqnarray}
\left\langle E\right\rangle &=&\left\langle \phi \left| \sum_{l} 
%TCIMACRO{\UNICODE[m]{0x127}}
%BeginExpansion
\rlap{\protect\rule[1.1ex]{.325em}{.1ex}}h%
%EndExpansion
\omega _{l}\hat{a}_{l}^{\dagger }\hat{a}_{l}\right| \phi \right\rangle 
\nonumber \\
&=&\sum_{l}%TCIMACRO{\UNICODE[m]{0x127}}
%BeginExpansion
\rlap{\protect\rule[1.1ex]{.325em}{.1ex}}h%
%EndExpansion
\omega _{l}\int W_{\phi }\left( \left\{ a_{l}\right\} \right) \left( \left|
a_{l}\right| ^{2}-\frac{1}{2}\right) d\func{Re}a_{l}d\func{Im}a_{l} 
\nonumber \\
&=&\sum_{l}%TCIMACRO{\UNICODE[m]{0x127}}
%BeginExpansion
\rlap{\protect\rule[1.1ex]{.325em}{.1ex}}h%
%EndExpansion
\omega _{l}\int \left| a_{l}\right| ^{2}\left( W_{\phi }\left( \left\{
a_{l}\right\} \right) -W_{0}\left( \left\{ a_{l}\right\} \right) \right) d%
\func{Re}a_{l}d\func{Im}a_{l},  \label{energy}
\end{eqnarray}
where $W_{\phi }\left( \left\{ a_{l}\right\} \right) $ is the state (Wigner
function) that in WW represents the HS state $\mid \phi \rangle $ and $W_{0}$
the vacuum Wigner function eq.$\left( \ref{1}\right) .$ The result is that
the \textit{normal ordering rule of HS corresponds to removing the ZPF energy%
} in practical calculations. Thus the vacuum Wigner function eq.$\left( \ref
{1}\right) $ plays a strange role in the WW representation. On the one hand
it should act on any charges. On the other hand it appears, via eq.$\left( 
\ref{energy}\right) ,$ as an inert stuff devoid of physical relevance,
except for its fluctuations. In the next section 4 we shall see that the ZPF
is essential in order to understand the WW representation of the EM field as
an explicit local model for the experiments involving ``entangled photon
pairs''.

\textbf{Energy flux of the radiation field. }In WW we may associate a linear
momentum 
\begin{equation}
\mathbf{p}=%TCIMACRO{\UNICODE[m]{0x127}}
%BeginExpansion
\rlap{\protect\rule[1.1ex]{.325em}{.1ex}}h%
%EndExpansion
\mathbf{k}_{l}\left| a_{l}\right| ^{2},  \label{p}
\end{equation}
to any plane wave, a quantity corresponding also to the energy flux crossing
a plane perpendicular to the wavevector. In WW the total energy flux due to
the vacuum field crossing any plane will be zero on the average due to
rotational invariance of the ZPF. That is 
\[
\left\langle \mathbf{p}_{0}\right\rangle =\sum_{l}\int 
%TCIMACRO{\UNICODE[m]{0x127}}
%BeginExpansion
\rlap{\protect\rule[1.1ex]{.325em}{.1ex}}h%
%EndExpansion
\mathbf{k}_{l}I_{l}W_{0}\left( \left\{ a_{l}\right\} \right) dI_{l}=0. 
\]
If we have a signal, that ideally we represent by a plane wave, the total
momentum of the signal plus ZPF will be only the one corresponding to the
signal. This contrast with what happens with the energy where the
(divergent) ZPF should be subtracted via a rather artificial procedure
derived from the ``normal ordering rule'' of HS.

These results provide a clue for a realistic interpretation of detection in
WW. Indeed the directional property of any radiation signal coming from a
source may explain why weak signals may be detected immersed in the very
strong ZPF always present. Of course the explanation is not required in HS
where the vacuum looks empty. A realistic interpretation of the correlation
experiments studied in the following should start from an interpretation of
photocounting, which is made in subsection 4.2 below.

\section{Experiments with entangled photon pairs}

\subsection{Spontaneous parametric down conversion in WW}

We are interested in experiments where the ``entangled photon pairs'' are
produced via spontaneous parametric down-conversion (SPDC). There is a
crystal possessing nonlinear electric susceptibility and a laser hitting on
the crystal, say on its left side. The process may be interpreted
classically (i.e. within Maxwell-Lorentz electrodynamics) \cite{Kaled}.
Indeed classical electrodynamics, and actually performed experiments, show
that if an intense laser beam and a macroscopic almost monochromatic light
beam, with frequencies $\omega _{L}$ and $\omega _{S}<\omega _{L}$
respectively, and different directions, both hit on an appropriate nonlinear
crystal, then three beams emerge from the opposite side of the crystal. That
is: 1) the incoming laser and 2) the incident macroscopic beam, both after
crossing the crystal in almost straight line and emerging with the same
frequencies, plus 3) another macroscopic beam with frequency $\omega
_{I}=\omega _{L}-\omega _{S}$ that emerges in a different direction.

The physics inside the crystal is as follows: the laser and the additional
incoming beam acts on charges (electrons) inside the nonlinear crystal
producing an oscillatory motion with frequency $\omega _{L}-\omega _{S},$
and the charges radiate preferently in a determined direction, giving rise
to the new exiting beam. If there are two beams incident on the crystal (in
addition to the laser) and the directions of the beams are appropriately
chosen (conjugate), then the outgoing beams have increased intensity because
they carry a part of the energy of the laser whose intensiy diminish.

A similar phenomenon appears if ``only the laser'' hits the crystal,
although in this case the emerging radiation is weak and it appears in the
form of a rainbow with several different colors. A plausible interpretation
of the latter phenomenon is that the ``vacuum'' radiation hitting on the
left side of the crystal, in addition to the laser, plays the role of the
incoming beams in the macroscopic experiment. This is the SPDC process
producing ``entangled photon pairs''.

In typical SPDC experiments \textit{two beams} with amplitudes $b_{s}$ and $%
b_{i}$ respectively, \textit{are selected}, via apertures and a lens system,
amongst the radiation emerging from the crystal, say on its right hand side
of the crystal. Working in the two-modes approximation, the amplitudes $%
b_{s} $ and $b_{s}$ of the selected outgoing beams are related to the
amplitudes $a_{s}$ and $a_{i}$ of the incoming vacuum modes as follows 
\begin{equation}
a_{s}\rightarrow b_{s}\equiv a_{s}+\gamma a_{i}^{*},a_{i}\rightarrow
b_{i}\equiv a_{i}+\gamma a_{s}^{*},\left| \gamma \right| <<1,  \label{pairs}
\end{equation}
where $\gamma $ is a complex number$.$ The modes with amplitudes $b_{s}$ and 
$b_{i}$ propagate with wavevectors \textbf{k}$_{s}$ and \textbf{k}$_{i}$
respectively. (It is common to label the outgoing beams ``signal'' and
``idler'' respectively, hence the subindices $s$ and $i$). The remaining
vacuum modes may be assumed to cross the crystal without any change. The
relevant result of the SPDC is the creation of new radiation represented by
the terms proportional to $\gamma $ in eqs.$\left( \ref{pairs}\right) .$ The
energy of that radiation comes from the laser, and the frequencies of signal
, $\omega _{s},$ idler , $\omega _{i},$ and laser, $\omega _{L},$ fulfil 
\begin{equation}
\omega _{L}=\omega _{s}+\omega _{i}.  \label{freq}
\end{equation}
In the HS formulation of quantum theory eq.$\left( \ref{freq}\right) $ is
commonly interpreted as energy conservation in the ``splitting of one laser
photon with energy $%TCIMACRO{\UNICODE[m]{0x127}}
%BeginExpansion
\rlap{\protect\rule[1.1ex]{.325em}{.1ex}}h%
%EndExpansion
\omega _{L}$ giving rise to two photons with energies $
%TCIMACRO{\UNICODE[m]{0x127}}
%BeginExpansion
\rlap{\protect\rule[1.1ex]{.325em}{.1ex}}h%
%EndExpansion
\omega _{s},$ and $%TCIMACRO{\UNICODE[m]{0x127}}
%BeginExpansion
\rlap{\protect\rule[1.1ex]{.325em}{.1ex}}h%
%EndExpansion
\omega _{i}".$ But in both the macroscopic experiment and the quantum WW
formalism eq.$\left( \ref{freq}\right) $ is plausibly interpreted as a
frequency matching condition, required for the creation of new radiation in
a nonlinear process involving charges of the crystal.

In order to describe the radiation emerging from the crystal within our
realistic WW interpretation we may still use the expansion eq.$\left( \ref{A}%
\right) $ except the substitution of $b_{s}$ and $b_{i}$ for $a_{s}$ and $%
a_{i}$, respectively. That is the expansion in plane waves of the radiation
field on the right hand side of the crystal will be 
\begin{eqnarray}
\mathbf{A}_{\phi }(\mathbf{x},t) &=&b_{s}\mathbf{\varepsilon }_{s}\exp
\left( i\mathbf{k}_{s}\mathbf{\cdot x-}i\omega _{s}t\right) +b_{i}\mathbf{%
\varepsilon }_{i}\exp \left( i\mathbf{k}_{i}\mathbf{\cdot x-}i\omega
_{i}t\right)  \nonumber \\
&&+\sum_{l\neq s,l\neq i}a_{l}\mathbf{\varepsilon }_{l}\exp \left( i\mathbf{k%
}_{l}\mathbf{\cdot x-}i\omega _{l}t\right) +\func{complex}conjugate 
\nonumber \\
&=&\mathbf{A}(\mathbf{x},t)+\gamma a_{i}^{*}\mathbf{\varepsilon }_{s}\exp
\left( i\mathbf{k}_{s}\mathbf{\cdot x-}i\omega _{s}t\right) +\gamma a_{s}^{*}%
\mathbf{\varepsilon }_{i}\exp \left( i\mathbf{k}_{i}\mathbf{\cdot x-}i\omega
_{i}t\right) +c.c.,  \label{B}
\end{eqnarray}
where $\mathbf{A}(\mathbf{x},t)$ is the radiation that would emerge from the
crystal with the pumping laser switched off. The probability distribution of
the modes $\left\{ a_{l}\right\} $ again consists of a product of Gaussians
similar to eq.$\left( \ref{1}\right) $.

In the correlation experiments the detectors are separated by macroscopic
distances. Therefore it is not appropriate to define the normal modes (plane
waves) within a too large normalization volume that includes everything,
e.g. both the source and the detectors. It is more convenient to use two
different normalization volumes, one aroung each detector. That is a
normalization volume with linear dimensions much smaller than the distances
amongst the detectors. As a consequence a different expansion in plane
waves, like eq.$\left( \ref{A}\right) ,$ shall be used for every detector.

\subsection{Experiment measuring the correlation between signal and idler}

In order to show how the WW representation may give rise to a local
realistic model, which is the main purpose of the present paper, I will work
a simple but illustrative example involving ``entangled photon pairs''
produced via SPDC. Of course in WW there are no photons, just (wave) fields,
but I use the current name for ease.

The experiment consists of measuring the single and coincidence detection
rates in two photocounters (Alice and Bob) appropriately placed in order to
detect the signal and idler radiation coming directly from the crystal. The
experiment exhibits a strong correlation characteristic of \textit{%
entanglement}. In the following I will analyze the experiment firstly within
HS, secondly in WW using the Weyl transform eq.$\left( \ref{Wtransform}%
\right) ,$ and finally within a local model derived from a realistic
interpretation of WW. The analysis of the experiment is relatively easy and
provides a guide for the more involved tests of the Bell inequalities to be
studied in section 6.

The quantities reported in typical correlation experiments are the single, $%
R_{A},R_{B},$ and coincidence, $R_{AB},$ detection rates measured by Alice
and Bob. For the interpretation it is more useful to study the single, $%
P_{A},P_{B},$ and coincidence, $P_{AB},$ detection probabilities within a
given detection time window. If $R_{0}$ is the rate of windows the following
relations hold $R_{A}=R_{0}P_{A},$ $R_{B}=R_{0}P_{B},$ $R_{AB}=R_{0}P_{AB}.$

\subsubsection{Analysis within the standard quantum HS formalism}

In the HS formalism the signal and idler beams producend in the nonlinear
crystal are represented by the operators (compare with eq.$\left( \ref{pairs}%
\right) $%
\begin{equation}
\hat{b}_{s}\equiv \ \hat{a}_{s}+\gamma \hat{a}_{i}^{\dagger },\hat{b}%
_{i}\equiv \hat{a}_{i}+\gamma \hat{a}_{s}^{\dagger }.  \label{qpairs}
\end{equation}
In order to calculate the detection probabilities in a time window it is
necessary to know the behaviour of the detectors (photocounters). In
practice the behaviour of detectors is \textit{postulated} via some rules
assumed plausible. In particular the single rates are supposed to be
proportional to the vacuum expectation values of products of the field
operators in normal order, that is 
\begin{equation}
P_{A}=K\left\langle 0\left| \hat{b}_{s}^{\dagger }\hat{b}_{s}\right|
0\right\rangle =\left| \gamma \right| ^{2},P_{B}=K\left\langle 0\left| \hat{b%
}_{i}^{\dagger }\hat{b}_{i}\right| 0\right\rangle =\left| \gamma \right|
^{2},K=1  \label{ra}
\end{equation}
The choice in eqs.$\left( \ref{ra}\right) $ is justified because $\hat{b}%
_{s}^{\dagger }\hat{b}_{s}$ $\left( \hat{b}_{i}^{\dagger }\hat{b}_{i}\right) 
$ is the ``photon number operator'' in the signal beam arriving at Alice
(the idler beam arriving at Bob). Then the constant K is the probability of
detection per window divided by the vacuum expectation of the operator
number of photons. In the following I will put $K=1$ for simplicity.

The choice for the coincidence probability $P_{AB}$ is consistent with the
election made for single detection. It is proposed that $P_{AB}$ equals the
expectation of the product of two-photon numbers, for Alice and Bob
respectively. However an ambiguity arises if the two creation (or
annihilation) operators do not commute with each other and the standard
choice is as follows 
\begin{equation}
P_{AB}=\frac{1}{2}\left\langle 0\left| \hat{b}_{s}^{\dagger }\hat{b}%
_{i}^{\dagger }\hat{b}_{i}\hat{b}_{s}\right| 0\right\rangle +\frac{1}{2}%
\left\langle 0\left| \hat{b}_{i}^{\dagger }\hat{b}_{s}^{\dagger }\hat{b}_{s}%
\hat{b}_{i}\right| 0\right\rangle =\left| \gamma \right| ^{2}+\left| \gamma
\right| ^{4}.  \label{rab}
\end{equation}
The result is straightforward taking eqs.$\left( \ref{qpairs}\right) $ into
account and they agree with experiments $($see e.g. Milonni) provided we
neglect terms of order $O\left( \left| \gamma \right| ^{4}\right) $ as is
usual practice because $\left| \gamma \right| ^{2}<<1.$

Eqs.$\left( \ref{ra}\right) $ and $\left( \ref{rab}\right) $ show a strong
correlation amongst the detecion rates, that is $R_{AB}\simeq R_{A}\simeq
R_{B},$ and therefore amongst detection probabilities $P_{AB}\simeq
P_{A}\simeq P_{B}$, which means that the probability of detection by Bob
(Alice), conditional to a detection by Alice (Bob), is unity. This is the
maximal correlation possible. Indeed if no correlation existed the
probability of joint detection would be much smaller than the single
detection$,$ that is $P_{AB}\simeq P_{A}P_{B}<<P_{A},P_{B}.$ Of course
non-idealities of the detectors make the correlation weaker in experimental
practice, that is $R_{AB}<R_{A},R_{AB}<R_{B}.$ Most non-idealities may be
taken into account introducing the efficiencies of the detectors so that we
should multiply $P_{A}$ and $P_{B}$ times $\eta _{1}<1$ and $\eta _{2}<1$
respectively, and $P_{AB}$ times the product $\eta _{1}\eta _{2}.$

From a fundamental point of view, however, the quantum predictions eqs.$%
\left( \ref{ra}\right) $\ and $\left( \ref{rab}\right) $\ cannot be correct
because in the ideal case the following inequality holds 
\begin{equation}
P_{A}-P_{AB}=P_{B}-P_{AB}=-\left| \gamma \right| ^{4}<0.  \label{ineq}
\end{equation}
which is absurd, probability theory demanding $P_{AB}\leq P_{A},P_{AB}\leq
P_{B}$. We should attribute the wrong prediction to the approximation of
order O$\left( \left| \gamma \right| ^{2}\right) $ involved in the
assumptions made for detection, eqs.$\left( \ref{ra}\right) $ and $\left( 
\ref{rab}\right) $. A similar problem will be appear in the WW treatment
that will be analyzed in section 5.3 below. Of course retaining terms of
order O$\left( \left| \gamma \right| ^{4}\right) $ in the applications of
SPDC is inconsistent, see eq.$\left( \ref{pairs}\right) .$

In order to understand more clearly the physics let us write eqs.$\left( \ref
{ra}\right) $ and $\left( \ref{rab}\right) $ in terms of the operators $\hat{%
a}$ and $\hat{a}^{\dagger }$ taking eqs.$\left( \ref{qpairs}\right) $ into
account, that is 
\begin{eqnarray}
P_{A} &=&\left\langle 0\left| \left( \hat{a}_{s}^{\dagger }+\gamma ^{*}\hat{a%
}_{i}\right) \left( \hat{a}_{s}+\gamma \hat{a}_{i}^{\dagger }\right) \right|
0\right\rangle =\left| \gamma \right| ^{2}\left\langle 0\left| \hat{a}_{i}%
\hat{a}_{i}^{\dagger }\right| 0\right\rangle =\left| \gamma \right| ^{2}, 
\nonumber \\
P_{AB}^{(1} &\equiv &\frac{1}{2}\left\langle 0\left| \hat{b}_{s}^{\dagger }%
\hat{b}_{i}^{\dagger }\hat{b}_{i}\hat{b}_{s}\right| 0\right\rangle =\frac{1}{%
2}\left\langle 0\left| \left( \hat{a}_{s}^{\dagger }+\gamma ^{*}\hat{a}%
_{i}\right) \left( \hat{a}_{i}^{\dagger }+\gamma ^{*}\hat{a}_{s}\right)
\left( \hat{a}_{i}+\gamma \hat{a}_{s}^{\dagger }\right) \left( \hat{a}%
_{s}+\gamma \hat{a}_{i}^{\dagger }\right) \right| 0\right\rangle  \nonumber
\\
&=&\frac{1}{2}\left| \gamma \right| ^{2}\left\langle 0\left| \hat{a}_{i}\hat{%
a}_{i}^{\dagger }\hat{a}_{i}\hat{a}_{i}^{\dagger }\right| 0\right\rangle +%
\frac{1}{2}\left| \gamma \right| ^{4}\left\langle 0\left| \hat{a}_{i}\hat{a}%
_{s}\hat{a}_{s}^{\dagger }\hat{a}_{i}^{\dagger }\right| 0\right\rangle =%
\frac{1}{2}\left( \left| \gamma \right| ^{2}+\left| \gamma \right|
^{4}\right) ,  \label{rab1}
\end{eqnarray}
and similar for $P_{B}=\left| \gamma \right| ^{2}$ , $P_{AB}^{(2}\equiv 
\frac{1}{2}\left\langle 0\left| \hat{b}_{i}^{\dagger }\hat{b}_{s}^{\dagger }%
\hat{b}_{s}\hat{b}_{i}\right| 0\right\rangle =\frac{1}{2}\left( \left|
\gamma \right| ^{2}+\left| \gamma \right| ^{4}\right) $ and 
\begin{equation}
P_{AB}=\left| \gamma \right| ^{2}+O(\left| \gamma \right| ^{4}).
\label{rab2}
\end{equation}

The following features may be realized. Firstly the normal ordering rule
prevents any contribution of order $\left| \gamma \right| ^{0},$ which fits
in the intuition because those terms could not derive from the signals
produced in the source. In fact they would appear even if the pumping laser
was switched off. For a similar reason there is no contribution of the
signal beam $\hat{a}_{s}$ (the idler beam $\hat{a}_{i}$) in the single rate
of Alice (Bob). On the other side the term $\hat{b}_{s}^{\dagger }\hat{b}%
_{i}^{\dagger }\hat{b}_{i}\hat{b}_{s}$ gives a contribution coming from $%
\hat{a}_{i}\hat{a}_{i}^{\dagger }\hat{a}_{i}\hat{a}_{i}^{\dagger },$ which
is a correlation between the \textit{inert} beam $\hat{a}_{i}$ going to Bob
and the \textit{active }beam $\hat{a}_{i}^{\dagger }$ going to Alice. There
is a similar term $\hat{a}_{s}\hat{a}_{s}^{\dagger }\hat{a}_{s}\hat{a}%
_{s}^{\dagger }$ coming from $\hat{b}_{i}^{\dagger }\hat{b}_{s}^{\dagger }%
\hat{b}_{s}\hat{b}_{i}$. These terms are essential for the strong
correlation.

I stress that the results of the experiment are not easy to understant
intuitively within HS. Indeed crucial elements are the commutation rules
between $\hat{a}$ and $\hat{a}^{\dagger }$ and the annihilation property of $%
\hat{a}^{\dagger }$. These are typically quantum properties inappropriate
for a realistic interpretation. In particular no picture emerges about the
nature of entanglement. As we shall see the understanding is improved going
from HS to the WW treatment and the interpretation provided in section 3.2.3.

\subsubsection{Analysis within the\textbf{\ Wigner representation, WW }}

The analysis of\textbf{\ }the experiment in WW should be got via the Weyl
transform applied to the detection assumptions eqs.$\left( \ref{ra}\right) $
and $\left( \ref{rab}\right) $ of HS. In any case these predictions should
agree with those of HS because WW and HS are different formalisms for the
same (quantum) theory of the EM field.

In order to go from HS to WW the expressions involving operators should
appear symmetrized, as required in eq.$\left( \ref{sym}\right) .$ In
particular for the intensity leading to the single detection probability $%
P_{A}$ we have 
\[
\hat{a}_{i}\hat{a}_{i}^{\dagger }=\frac{1}{2}\left( \hat{a}_{i}\hat{a}%
_{i}^{\dagger }+\hat{a}_{i}^{\dagger }\hat{a}_{i}\right) +\frac{1}{2}\left( 
\hat{a}_{i}\hat{a}_{i}^{\dagger }-\hat{a}_{i}^{\dagger }\hat{a}_{i}\right) =%
\frac{1}{2}\left( \hat{a}_{i}\hat{a}_{i}^{\dagger }+\hat{a}_{i}^{\dagger }%
\hat{a}_{i}\right) +\frac{1}{2}. 
\]
Hence eq.$\left( \ref{sym}\right) $ gives the WW expression

\begin{equation}
P_{A}=\left| \gamma \right| ^{2}\left\langle \left| a_{i}\right| ^{2}+\frac{1%
}{2}\right\rangle =\left| \gamma \right| ^{2},  \label{0p}
\end{equation}
where $\left\langle {}\right\rangle $ means integration with respect to the
amplitudes of all modes with the distribution $W_{0}$ eq.$\left( \ref{1}%
\right) ,$ which leads to $\left\langle \left| a_{i}\right|
^{2}\right\rangle =\frac{1}{2}$ (the integral of the remaining modes, with $%
j\neq 1$ contribute a factor unity). And similar for P$_{B}$, that is 
\begin{equation}
P_{A}=2\left| \gamma \right| ^{2}\left\langle \left| a_{i}\right|
^{2}\right\rangle =\left| \gamma \right| ^{2},P_{B}=2\left| \gamma \right|
^{2}\left\langle \left| a_{s}\right| ^{2}\right\rangle =\left| \gamma
\right| ^{2},  \label{PWW}
\end{equation}
in agreement with HS.

In order to get the Weyl transform for the coincidence probability $P_{AB}$
I will start writing eq.$\left( \ref{rab1}\right) $ in terms of symmetrized
products of operators. Then we shall perform the Weyl transform taking eq.$%
\left( \ref{sym}\right) $ into account. For the first term of $\left( \ref
{rab1}\right) $ we get 
\begin{eqnarray*}
\left\langle 0\left| \hat{a}_{i}\hat{a}_{i}^{\dagger }\hat{a}_{i}\hat{a}%
_{i}^{\dagger }\right| 0\right\rangle &=&\left\langle 0\left| (\hat{a}_{i}%
\hat{a}_{i}^{\dagger }+\hat{a}_{i}^{\dagger }\hat{a}_{i})(\hat{a}_{i}\hat{a}%
_{i}^{\dagger }+\hat{a}_{i}^{\dagger }\hat{a}_{i})\right| 0\right\rangle \\
&=&\left\langle a_{i}a_{i}^{*}a_{i}a_{i}^{*}\right\rangle =\left\langle
\left| a_{i}\right| ^{4}\right\rangle ,
\end{eqnarray*}
where the first equality is correct because the terms added in the right
sides do not contribute and the second equality is the Weyl transform
leading from HS to WW expressions. Similarly the second term of eq.$\left( 
\ref{rab1}\right) $ leads to 
\begin{eqnarray*}
\left\langle 0\left| \hat{a}_{i}\hat{a}_{s}\hat{a}_{s}^{\dagger }\hat{a}%
_{i}^{\dagger }\right| 0\right\rangle &=&\left\langle 0\left| \hat{a}_{s}%
\hat{a}_{s}^{\dagger }\hat{a}_{i}\hat{a}_{i}^{\dagger }\right| 0\right\rangle
\\
&=&\left\langle 0\left| (\hat{a}_{s}\hat{a}_{s}^{\dagger }+\hat{a}%
_{s}^{\dagger }\hat{a}_{s})(\hat{a}_{i}\hat{a}_{i}^{\dagger }+\hat{a}%
_{i}^{\dagger }\hat{a}_{i})\right| 0\right\rangle =\left\langle \left|
a_{s}\right| ^{2}\left| a_{i}\right| ^{2}\right\rangle .
\end{eqnarray*}
After some algebra we get for $P_{AB}=P_{AB}^{(1}+P_{AB}^{(2}$%
\begin{eqnarray}
P_{AB} &=&\left| \gamma \right| ^{2}\left[ \left( \left\langle \left|
a_{s}\right| ^{4}\right\rangle +\left\langle \left| a_{i}\right|
^{4}\right\rangle \right) +2\left( \left\langle \left| a_{s}\right|
^{2}\right\rangle ^{2}+\left\langle \left| a_{i}\right| ^{2}\right\rangle
^{2}\right) \right]  \nonumber \\
&&+4\left| \gamma \right| ^{4}\left\langle \left| a_{s}\right|
^{2}\right\rangle \left\langle \left| a_{i}\right| ^{2}\right\rangle =\left|
\gamma \right| ^{2}+O(\left| \gamma \right| ^{4}),  \label{PAB}
\end{eqnarray}
which reproduces the quantum result eq.$\left( \ref{rab}\right) $ obtained
in HS, as expected.

\subsubsection{What is ``photon'' entanglement?}

In WW there are no photons, radiation appears as continuous (waves). Hence
entanglement should correspond to some correlation between fields. In fact
field intensities flutuate and coincidence detection should be more probable
when a positive big fluctuation arrives simultaneously to both Alice and
Bob. With reference to eq.$\left( \ref{pairs}\right) $ we may realize that
entanglement appears as a correlation between two light beams, 1) the ZPF
radiation with amplitude $a_{s}$ that entered the nonlinear crystal, crossed
it and emerged going to Alice and 2) the radiation with amplitude $\gamma
a_{s}^{*}$ created in the crystal and added to the ZPF mode with amplitude $%
a_{i}$ going to Bob. And similarly for the correlation between $a_{i}$ going
to Bob and $\gamma a_{i}^{*}$ going to Alice. That is in our analysis with
WW \textit{entanglement is a correlation between a signal }(either $\gamma
a_{s}^{*}$ or $\gamma a_{i}^{*})$ \textit{and a part of the vacuum fields ZPF%
} (either $a_{s}$ or $a_{i})$. It is nonclassical because the vacuum field
ZPF is a quantum feature. Furthermore the existence of ZPF is the essential
difference between the classical and the electromagnetic field quantized via
the Wigner (WW) representation. This interpretation will be more clear in
the LR models that follow.

\section{The quantized field in WW as a local realistic model}

\subsection{Introduction}

In the following I propose a realistic (or ``hidden variables'') local
interpretation of the \textit{quantum} EM field as it appears in the WW
formalism. I include possible interaction with macroscopic bodies. As said
above the resulting formalism of the quantum field is identical to classical
electrodynamics, \textit{except} for the assumption that there is a random
radiation with distribution eq.$\left( \ref{1}\right) $ filling space. In
particular the propagation of the radiation is causal in the sense of
relativity theory and the interaction of the field with macroscopic objects
fits in classical electrodynamics. Therefore our interpretation of the
quantum WW formalism provides in principle a local realistic (LR) model for
experiments within a restricted domain. The model cannot be generalized to
phenomena involving either quantum fields other than electromagnetism, or
microscopic charges like electrons or atoms, because this would requiere a
quantum theory of those fields or particles formulated in a way similar to
WW for the EM field, which is not available.

For the sake of clarity I point out that \textit{the equivalence between the
standard (HS) formulation and WW refers to the formalism, but not
necessarily to the connection with empirical facts}. In particular
``photons'' are well defined in the \textit{mathematical }Hilbert space, but
they do not correspond to concepts of the reality. I hope this was clear
enough in section 2.2. Therefore in order to get a local realistic model
resting on WW I shall not assume that the measurement theory of standard
(HS) quantum theory should be translated to WW. \textit{Actually in a local
model based on WW we must introduce an appropriate substitute for the HS
measurement theory. }In the treament of optical correlation experiments the
detections are the only measurement required. Then the assumptions required
for the connection of the WW formalism with experiments consists of
assumptions about the response of the detectors (photocounters) to the
arriving radiation. That is the probability p(I) of a photocount as a
function of the intensity, I, integrated within a given time window. I
stress that in a local realistic model, as in classical physics, I do not
propose a general measurement theory, but particular rules for specific
experimental set ups.

The local models proposed below will allow the interpretation of experiments
with ``entangled photon pairs'', which involve the EM field plus macroscopic
set-ups like laser beams, nonlinear crystals, photocounters and
beam-splitters, plus other standard devices in optical experiments. (I have
written ``entangled photon pairs'' in order to use a common language but I
point out again that the EM radiation consists of just waves in our
realistic interpretation of WW). The question whether the models agrees with
relevant experiments, in particular optical tests of the Bell inequalities
will be treated in section 5.

I shall mention another apparent difficulty in our approach. In fact a
realistic interpretation of optical correlation experiments does not fit in
the few modes approximation used in section 3, that is a few plane waves
extended over a large normalization volume as in eq.$\left( \ref{A}\right) .$
In actual experiments people selects narrow beams which are quite different
from single field modes, Indeed a radiation beam should be represented by a
linear combination of many plane waves.

For a single mode the amplitude is Gaussian with zero mean whence the
distribution of the intensity $I_{j}=\left| a_{j}\right| ^{2}$ is an
exponential function. For all modes we may write the intensity as follows,
taking eq.$\left( \ref{1}\right) $ into account, 
\begin{equation}
W_{0}\left( \left\{ I_{j}\right\} \right) =\prod_{j}2\exp \left(
-2I_{j}\right) \text{ with }\int W_{0}\left( \left\{ I_{j}\right\} \right)
\prod_{j}dI_{j}=1.  \label{1a}
\end{equation}
The total intensity of a beam consisting of many modes (plane waves) will be
the square modulus of the total amplitude, which corresponds to the sum of
the amplitudes of the modes. These amplitudes are random variables with
similar means and statistically independent of each other. Then it is
plausible to assume that the total intensity of the beam is a Gaussian
random variable with a narrow distribution around the mean, whence I propose
that the total intensity $W_{beam}\left( I\right) $ of a radiation beam is
the following Gaussian 
\begin{equation}
W_{beam}\left( I\right) =\sqrt{\frac{\alpha }{\pi }}\exp \left[ -\alpha
\left( I-1/2\right) ^{2}\right] .  \label{beam}
\end{equation}
The parameter $\alpha $ will depend on the number of plane waves involved,
which on turn is determined by the apertures defining the beam in the
experiment. For the intensities I will choose arbitrary units, so that I
define that the mean of the beam is 1/2 in our (arbitrary dimensionless)
units. The parameter $\alpha $ is assumed large so that the distribution eq.$%
\left( \ref{beam}\right) $ is narrow as said above.

In spite of what I have said, will use the few modes approximation and
therefore the distribution eq.$\left( \ref{1a}\right) $ in the following two
sections 4.2 and 4.3, in order to stress the analogy of the model with the
WW representation. In section 4.4 I will change to a more realistic many
modes treatment using the distribution eq.$\left( \ref{beam}\right) .$

\subsection{Photodetection in the presence of a real vacuum field}

As said at the begining of section 3.2 the quantities measured in the
experiments are detection rates but it is convenient to consider instead the
detection probabilities within appropriate time windows in order to
interpret the results. A detection rate $R$ will be the product of the
detection probability, $P$, within a time window times the rate of windows, $%
R_{0}.$ In the standard HS formalism it is common to propose some plausible
rules for detection probabilities as was commented on section 2.1. Typically
people assumes in HS that the detection probability is proportional to the
number of photons reaching the detector in a time window, which plausibly
becomes proportional to the intensity of the radiation, $I$, in the WW
formalism. However as said in section 4.1 that assumption is incorrect.

As commented on the final paragraphs of section 2.3, it is more convenient
in our approach to relate detection with the directional energy flux $%
\mathbf{p}$ defined in eq.$\left( \ref{p}\right) .$ In order to fit in the
common use of the radiation intensity, as in section 3, we define a
directional, or vector, intensity of the radiation mode $l$ by 
\begin{equation}
\vec{I}_{l}\equiv %TCIMACRO{\UNICODE[m]{0x127}}
%BeginExpansion
\rlap{\protect\rule[1.1ex]{.325em}{.1ex}}h%
%EndExpansion
\mathbf{k}_{l}/(c\omega _{l})\left| a_{l}\right| ^{2}.  \label{I0}
\end{equation}
(The vector intensity will be distinguished from the usual scalar intensity $%
I_{l}=\left| a_{l}\right| ^{2}$, by an arrow above the letter as in eq.$%
\left( \ref{I0}\right) ).$

Now I shall define a detection theory, that is the probability of a
photocount within a time window, as a function of the intensity arriving at
the detector. For the sake of clarity in the exposition I will tentatively
assume that the detection probability is proportional to the clean enegy
flux arriving to the detector in the appropriate direction. That is 
\begin{equation}
P\varpropto \sum_{l}\mathbf{u\cdot k}_{l}\left| a_{l}\right| ^{2},
\label{II}
\end{equation}
where $\mathbf{u}$ is a unit vector in a fixed direction, usually pointing
from the source of radiation towards the detector. However as said above and
we shall see more clearly in section 4.4 that simple proportionality between
intensity, I, and detection probability, P, is untenable in local realistic
models. There I will propose adequate modifications.

The detector is typically sensitive to a small range of frequencies and for
that range the vector intensity $\vec{I}_{l}$ may be taken to be
proportional to the energy flux, $\vec{I}_{l}\mathbf{=} 
%TCIMACRO{\UNICODE[m]{0x127}}
%BeginExpansion
\rlap{\protect\rule[1.1ex]{.325em}{.1ex}}h%
%EndExpansion
\mathbf{k}_{l}/(c\omega _{0})\left| a_{l}\right| ^{2}$. That is we may
substitute a single frequency $\omega _{0}$ for any frequencies $\omega _{l}$
in eq.$\left( \ref{I0}\right) ,$ within the range of sensitivity. In
summary, the probability of a detection event, within a given time window,
is here assumed to be 
\begin{equation}
P=KI,I=\mathbf{u\cdot }\sum_{l}\vec{I}_{l}  \label{I2}
\end{equation}
where $I$ is the (usual, scalar) intensity actually reaching the detector
averaged within a time window. As $I$ is a random variable we should average
also over its probability distribution $W\left( I\right) $, that is the
following should be substituted for eq.$\left( \ref{I2}\right) $ 
\begin{equation}
P=K\left\langle I\right\rangle \equiv K\int_{0}^{\infty }IW\left( I\right)
dI,  \label{2I}
\end{equation}
where W(I) is given by eq.$\left( \ref{1a}\right) $. This assumption is a
generalization of the hypothesis eq.$\left( \ref{PWW}\right) $ for the
detection probability as a function of the intensity which I used in the
study via WW of correlation experiments in section 3.2.2. I point out that
eqs.$\left( \ref{I2}\right) $ and $\left( \ref{2I}\right) $ include the
radiation due to the vacuum field ZPF.

Now we may attain a picture of detection in the presence of ZPF because in
the WW formalism the vacuum consists of a random radiation, see eq.$\left( 
\ref{1}\right) .$ That is we must explain why \textit{weak signals} may be
discriminated from the possibly \textit{stronger vacuum radiation} ZPF. Let
us assume that the detector receives radiation with vector intensity $\vec{I}%
_{s}$ produced in the source. The detector will be also hit by the ZPF
present everywhere. Then the effective scalar intensity $I$ arriving at the
detector will be the projection of the sum of all vector intensities times
the unit vector $\vec{s}=$ $I_{s}^{-1}\vec{I}_{s}$ in the direction of $\vec{%
I}_{s},$ which is also the direction determined by the position of the
detector with respect to the source (say the nonlinear crystal in parametric
down-conversion), that is 
\begin{equation}
I=\vec{I}\cdot \vec{s}=I_{s}+\sum_{l}\vec{I}_{l}\cdot \vec{s}\equiv
I_{s}+Z_{s},  \label{I3}
\end{equation}
where $\vec{I}_{l}$ was given in eq.$\left( \ref{I0}\right) $, $W_{0}$ in eq.%
$\left( \ref{1a}\right) $ and $Z_{s}$ comes from the (vacuum) ZPF. Here we
are assuming that $\vec{I}$ is not a part of the ZPF, in the next section
4.3 we will study the case where $\vec{I}$ does belong to the ZPF.

The average contribution of the ZPF, $\left\langle Z_{s}\right\rangle ,$ is
nil due to the rotational invariance of the ZPF. Hence, assuming eq.$\left( 
\ref{2I}\right) $, the detection probability will be $P=K\left\langle
I_{s}\right\rangle $ which may explain why a weak signal may be detected in
the presence of strong ZPF. However the fluctuations of the ZPF may be
relevant, as will be discussed in the next subsection 4.3.

\subsection{Local model for the signal-idler correlation experiment}

The first application of our approach will be a local model for the
signal-idler correlation experiment. The model may be seen as a realistic
interpretation of the WW approach handled in section 3.2.2. Our aim is to
derive eqs.$\left( \ref{PWW}\right) $ and $\left( \ref{PAB}\right) $ taking
the radiation produced in the SPDC process, eq. $\left( \ref{pairs}\right) ,$%
into account. The radiation is assumed to evolve according to classical
electrodynamics, see section 3.3. Our model will be presented here assuming
the detection probability eq.$\left( \ref{2I}\right) ,$ and required
modifications will be studied in section 4.4.

Our aim here is to derive the single and coincidence detection probabilities
in the signal-idler correlation experiment described in section 3.2 treating
the problem within classical electrodynamics, but assuming the existence of
the random field ZPF. The results shall agree with the quantum predictions
eqs.$\left( \ref{PWW}\right) $ and $\left( \ref{PAB}\right) .$ Alice\'{}s
detector receives an intensity, $I_{A}$ derived from the field amplitude eq.$%
\left( \ref{pairs}\right) ,$ that is 
\begin{equation}
I_{A}=\left| a_{s}+\gamma a_{i}^{*}\right| ^{2}=\left| a_{s}\right|
^{2}+\left| \gamma \right| ^{2}\left| a_{i}\right| ^{2}=I_{s}+\left| \gamma
\right| ^{2}I_{i},  \label{Alice}
\end{equation}
where I ignore a cross term $\left| \gamma \right| 2\func{Re}\left(
a_{s}a_{i}^{*}\right) $ that does not contribute to detection, having zero
mean because $a_{s}$ and $a_{i}^{*}$ are uncorrelated. Similarly the
intensity arriving at Bob\'{}s detector will be 
\begin{equation}
I_{B}=\left| a_{i}+\gamma a_{s}^{*}\right| ^{2}=I_{i}+\left| \gamma \right|
^{2}I_{s}.  \label{Bob}
\end{equation}
If no ZPF was present then the single detection probabilities in a time
window would equal the mean values of (the random variables) $I_{A}$ and $%
I_{B}$ times some constant $K$ depending on the duration of the time window
and the properties of the detector. Here I shall take $K=1$ whence the
detection probabilities would be the averages weighted by the distribution
eq.$\left( \ref{1a}\right) $, that we should write 
\begin{equation}
P_{A}^{c}=\left\langle I_{s}\right\rangle +\left\langle \left| \gamma
\right| ^{2}I_{i}\right\rangle =\frac{1}{2}+\frac{1}{2}\left| \gamma \right|
^{2},  \label{M0}
\end{equation}
and similar por $P_{B}^{c}.$ The upper index \textit{c} stands for
``classical''.

The result eq.$\left( \ref{M0}\right) $ does not agree with the experiment.
In particular it predicts some detection rate even if $\gamma =0$ which
would correspond to the pumping laser switched off. I claim that it is
because the ZPF has not been taken appropriately into account. In fact the
radiation mode with intensity $I_{s}$ does belong to the ZPF, as shown in
our derivation of eqs.$\left( \ref{pairs}\right) .$ Also we have ignored the
remaining ZPF arriving at Alice detector. The correct approach is to write
the total vector intensity arriving at Alice , that is (see eq.$\left( \ref
{I3}\right) $) 
\[
\vec{I}_{A}=\left| \gamma \right| ^{2}I_{i}\vec{s}+\sum_{l\neq s}\vec{I}_{l}+%
\vec{I}_{s}=\left| \gamma \right| ^{2}I_{i}\vec{s}+\sum_{l}\vec{I}_{l}, 
\]
where $\vec{s}$ was defined in section 4.2 and $\sum_{l}\vec{I}_{l}$ is a
form of representing the ZPF vector intensity arriving at the detector,
taking into account that $I_{s}$ is also a part of the ZPF. Indeed the
radiation with intensity $I_{s}$ arrived at the nonlinear crystal, crossed
it and after exiting it was sent to the detector, see section 3.1. Similarly
for the vector intensity arriving at Bob, that is 
\[
\vec{I}_{B}=\left| \gamma \right| ^{2}I_{s}\vec{i}+\sum_{j\neq i}\vec{I}_{j}+%
\vec{I}_{i}=\left| \gamma \right| ^{2}I_{s}\vec{i}+\sum_{j}\vec{I}_{j}, 
\]
I assume that the intensity relevant for detection by Alice is the
projection of the total incoming vector intensity on the direction of $\vec{s%
}.$

For clarity I define two new random variables, $Y_{A}$ and $Y_{B},$
representing (scalar) intensities of ZPF arriving at Alice and Bob detectors
respectively. That is 
\begin{equation}
Y_{A}\equiv \sum_{l\neq s}\vec{I}_{l}\cdot \vec{s},Y_{B}\equiv \sum_{j\neq i}%
\vec{I}_{j}\cdot \vec{i},  \label{Z}
\end{equation}
which are statistically independent of the intensities produced in the
source, $I_{s}$ and $\vec{I}_{i}$ except for the condition that the averages
fulfil 
\begin{equation}
\left\langle Y_{A}+I_{s}\right\rangle =0,\left\langle
Y_{B}+I_{i}\right\rangle =0.  \nonumber
\end{equation}
Hence it is convenient to introduce new variables defining 
\begin{equation}
Y_{A}=-\left\langle I_{s}\right\rangle +Z_{A},Y_{B}=-\left\langle
I_{i}\right\rangle +Z_{B}.  \label{Z1}
\end{equation}
The random variables $Z_{A}$ and $Z_{B}$ may be assumed Gaussian because
they are a sum of many independent variables with similar statistical
properties. They are uncorrelated and their probability distributions are
assumed equal, that is 
\begin{equation}
\left\langle Z_{A}^{n}\right\rangle =\left\langle Z_{B}^{n}\right\rangle
=\left\langle Z^{n}\right\rangle ,\smallskip \left\langle
Z_{A}^{m}Z_{B}^{n}\right\rangle =\smallskip \left\langle
Z_{A}^{m}\right\rangle \smallskip \left\langle Z_{B}^{n}\right\rangle
,\left\langle Z\right\rangle =0,  \label{M3}
\end{equation}
for any natural numbers $m,n$. The average $\left\langle Z\right\rangle $ is
nil by rotational symmetry of the vacuum radiation and we may assume a
Gaussian distribution, that is 
\begin{equation}
W\left( Z\right) =\sqrt{\frac{\beta }{\pi }}\exp \left[ -\beta Z^{2}\right] ,
\label{WZ}
\end{equation}
either for Z$_{A}$ or Z$_{B}$. Thus the intensities arriving at Alice and
Bob detectors may be written, respectively 
\begin{eqnarray}
I_{A} &=&I_{s}+\left| \gamma \right| ^{2}I_{i}+Y_{A}=I_{s}+\left| \gamma
\right| ^{2}I_{i}-\left\langle I_{s}\right\rangle +Z_{A},  \label{M2} \\
I_{B} &=&I_{i}+\left| \gamma \right| ^{2}I_{s}+Y_{B}=I_{i}+\left| \gamma
\right| ^{2}I_{s}-\left\langle I_{i}\right\rangle +Z_{B}.  \nonumber
\end{eqnarray}

The single detection probability by Alice would be an average (weighted by
the distribution eq.$\left( \ref{1a}\right) $ and the distribution of ZPF)
of the effective intensity $I_{A}$ eq.$\left( \ref{M2}\right) $ arriving at
Alice detector, that is 
\begin{equation}
P_{A}=\left\langle I_{A}\right\rangle =\frac{1}{2}\left| \gamma \right| ^{2},
\label{PA}
\end{equation}
where the nil value of $\left\langle Z_{A}\right\rangle $ has been taken
into account. A similar result is obtained for $P_{B}.$ There is agreement
with the WW eq.$\left( \ref{PWW}\right) $ except for the global factor $1/2$
, a discrepancy that derives from a different choice of units. In fact in HS
the unit was one photon or, equivalently, an energy $
%TCIMACRO{\UNICODE[m]{0x127}}
%BeginExpansion
\rlap{\protect\rule[1.1ex]{.325em}{.1ex}}h%
%EndExpansion
\omega ,$see eq.$\left( \ref{intensity}\right) .$ Here the units are related
to the distribution eq.$\left( \ref{1}\right) $ putting $\left\langle \left|
a_{l}\right| ^{2}\right\rangle =1/2.$

The coincidence detection probability is the average of a product of two
intensities, that is 
\begin{eqnarray}
P_{AB} &=&\left\langle I_{A}I_{B}\right\rangle =\left\langle \left(
I_{s}+\left| \gamma \right| ^{2}I_{i}+Y_{A}\right) \left( I_{i}+\left|
\gamma \right| ^{2}I_{s}+Y_{B}\right) \right\rangle  \label{AB} \\
&=&\left\langle \left( I_{s}+\left| \gamma \right| ^{2}I_{i}\right) \left(
I_{i}+\left| \gamma \right| ^{2}I_{s}\right) \right\rangle +\left\langle
\left( I_{s}+\left| \gamma \right| ^{2}I_{i}\right) \right\rangle
\left\langle Y_{B}\right\rangle  \nonumber \\
&&+\left\langle Y_{A}\right\rangle \left\langle \left( I_{i}+\left| \gamma
\right| ^{2}I_{s}\right) \right\rangle +\left\langle Y_{A}\right\rangle
\left\langle Y_{B}\right\rangle  \nonumber \\
&=&\left\langle \left( I_{s}+\left| \gamma \right| ^{2}I_{i}\right) \left(
I_{i}+\left| \gamma \right| ^{2}I_{s}\right) \right\rangle +\left\langle
\left( I_{s}+\left| \gamma \right| ^{2}I_{i}\right) \right\rangle
\left\langle -I_{i}\right\rangle  \nonumber \\
&&+\left\langle -I_{s}\right\rangle \left\langle \left( I_{i}+\left| \gamma
\right| ^{2}I_{s}\right) \right\rangle +\left\langle -I_{s}\right\rangle
\left\langle -I_{i}\right\rangle  \nonumber \\
&=&\left| \gamma \right| ^{4}\left\langle I_{s}\right\rangle \left\langle
I_{i}\right\rangle +\left| \gamma \right| ^{2}\left( \left\langle
I_{s}^{2}\right\rangle +\left\langle I_{i}^{2}\right\rangle \right) , 
\nonumber
\end{eqnarray}
where we have taken into account into account eqs.$\left( \ref{M2}\right) $
and the fact that all intensities involved are uncorrelated except the
obvious pairs $\left\langle I_{s}^{2}\right\rangle \neq \left\langle
I_{s}\right\rangle \left\langle I_{s}\right\rangle $ and $\left\langle
I_{i}^{2}\right\rangle \neq \left\langle I_{i}\right\rangle \left\langle
I_{i}\right\rangle .$ 
\begin{equation}
P_{AB}=\left| \gamma \right| ^{4}\left\langle I_{s}\right\rangle
\left\langle I_{i}\right\rangle +\left| \gamma \right| ^{2}\left(
\left\langle I_{s}^{2}\right\rangle -\left\langle I_{s}\right\rangle
^{2}\right) +\left| \gamma \right| ^{2}\left( \left\langle
I_{i}^{2}\right\rangle -\left\langle I_{i}\right\rangle ^{2}\right) =\frac{1%
}{2}\left| \gamma \right| ^{2}+\left| \gamma \right| ^{4},  \label{Q1}
\end{equation}
in agreement with the standard quantum result eq.$\left( \ref{PAB}\right) $
except for the factor 1/2. Hence we get, to order O$\left( \left| \gamma
\right| ^{2}\right) ,$ the standard quantum result eqs.$\left( \ref{ra}%
\right) $ and $\left( \ref{rab}\right) .$

In summary this subsection has shown that the WW formalism for the quantum
EM field may be interpreted as a local realistic (or hidden variables) model
of the (signal-idler) correlation experiment of two entangled beams produced
via spontaneous parametric down-conversion,\textit{\ modulo a treatment of
the detection postulate, eq.}$\left( \ref{2I}\right) ,$ which will be shown
unphysical in the next subsection 4.4.

\subsection{Improved model with more realistic detection postulates}

The need of changing the detection rule eq.$\left( \ref{2I}\right) $ arises
because it violates constraints of probability theory, whence the model
cannot be labeled realistic. Indeed any correct single detection probability
in a realistic model should be of the form 
\begin{equation}
P=\left\langle p\left( I\right) \right\rangle \equiv \int p\left( I\right)
W\left( I\right) dI,  \label{51}
\end{equation}
where $p\left( I\right) $ is the probability conditional to a total
intensity $I$ arriving at the detector within a time window. Then we should
average with respect to the possible values of $I$ with the probability
distribution $W\left( I\right) .$ In fact the function $p\left( I\right) $,
being a probability, is constrained by 
\begin{equation}
0\leq p\left( I\right) \leq 1,\text{ for all }I.  \label{52}
\end{equation}
These conditions do not hold for the detection probabilities eq.$\left( \ref
{2I}\right) ,$ which are equivalent to assuming $p\left( I\right) =KI$ that
violates eq.$\left( \ref{52}\right) $ for large intensity, that is whenever $%
I>1/K.$ Therefore the \textit{WW formalism, with the detection rules }%
translated from the HS formalism via the Weyl transform, as in sections
3.2.2 and 4.3,\textit{\ does not provide local realistic models directly. }%
In fact\textit{\ }some modifications are needed.

Another reason for the change is the use of a many-modes treatment required
for the distribution of intensities eq.$\left( \ref{beam}\right) .$ We shall
take into account that the radiation arriving at Alice consists of 3 types
of radiation modes (usually plane waves). Firstly there are modes that
entered the nonlinear crystal, crossed it and arrived to Alice. They
correspond to the single mode labelled $I_{s}$ in section 4.3, a label that
I shall use for the whole beam in the following. Secondly the amplitudes
corresponding to the previous radiation created in the crystal, named $%
\left| \gamma \right| ^{2}I_{i},$ a label that I shall use also for the
beam. Finnaly there is the radiation coming from the ZPF that I will name $%
Y_{A}$ as in the previous section 3.3, see eq.$\left( \ref{M2}\right) .$ The
latter may be written in terms of the full radiation arriving at Alice, that
is $Y_{A}=Z_{A}-1/2,$ see eq.$\left( \ref{Z1}\right) .$ Similarly for the
radiation arriving to Bob. Taking eqs.$\left( \ref{beam}\right) $ and $%
\left( \ref{WZ}\right) $ into account the distribution of the radiation will
be 
\begin{equation}
W=\frac{\alpha }{\pi }\exp \left[ -\alpha \left( I_{s}-1/2\right)
^{2}-\alpha \left( I_{i}-1/2\right) ^{2}\right] \sqrt{\frac{\beta }{\pi }}%
\exp \left[ -\beta Z_{A}^{2}\right] .  \label{WIZ}
\end{equation}
where I assume that Gaussians are a fair approximation for the two beams.

In order to illustrate the changes produced when both physical beams and
realistic detection assmptions are used, I will work a simple local model
for the experiment described in section 3.2. It is plausible that the
probability $p\left( I\right) $ increases with the intensity, $I$. We might
choose a function with a sharp discontinouity, that is a threshold model
where the detection probability is nil for $I<I_{0},$ and a constant for $%
I<I_{0}.$ However here I shall choose a smooth function, which is more
illustrative and allows easier computations. That is the following detection
function 
\begin{equation}
p\left( I\right) =1-\exp \left( -kI\right) ,k>0,  \label{4a}
\end{equation}
$k$ being a parameter of the model. It is easy to see that eq.$\left( \ref
{4a}\right) $ is equivalent to eq.$\left( \ref{2I}\right) $ for small $k$
but the constraints eq.$\left( \ref{52}\right) $ are never violated.

Then, neglecting the ZPF contribution, Z$_{A}$ and\ Z$_{B}$ eq.$\left( \ref
{M3}\right) ,$the single detection probabilities for the experiment
presented in section 3.2 are

\begin{eqnarray}
P_{A} &=&\int WdI_{s}dI_{i}\left[ 1-\exp \left( -kI_{s}-k\left| \gamma
\right| ^{2}I_{i}+k/2\right) \right] .  \nonumber \\
&=&1-\exp \left[ -\frac{1}{2}k\left| \gamma \right| ^{2}-\frac{k^{2}}{%
4\alpha }(1+\left| \gamma \right| ^{4})\right] =P_{B},  \label{13}
\end{eqnarray}
where W is given in eq.$\left( \ref{WIZ}\right) ,$ ignoring Z$_{A}$. For the
proof that $P_{A}=P_{B}$ it is enough to exchange $I_{s}\leftrightarrow
I_{i}.$ The coincidence detection probability is

\begin{eqnarray}
P_{AB} &=&\int WdI_{s}dI_{i}\left[ 1-\exp \left( -kI_{s}-k\left| \gamma
\right| ^{2}I_{i}+k/2\right) \right]  \nonumber \\
&&\times \left[ 1-\exp \left( -kI_{i}-k\left| \gamma \right|
^{2}I_{s}+k/2\right) \right] =P_{A}+P_{B}-1  \nonumber \\
&&+\int WdI_{s}dI_{i}\exp \left[ -k\left( I_{s}+I_{i}\right) \left( 1+\left|
\gamma \right| ^{2}\right) +k\right]  \nonumber \\
&=&1-2\exp \left[ -\frac{1}{2}k\left| \gamma \right| ^{2}-\frac{k^{2}}{%
4\alpha }(1+\left| \gamma \right| ^{4})\right]  \label{15} \\
&&+\exp \left[ -k\left( 1+\left| \gamma \right| ^{2}\right) -\frac{k^{2}}{%
2\alpha }\left( 1+\left| \gamma \right| ^{2}\right) ^{2}\right] .  \nonumber
\end{eqnarray}
Both single and coincidence detection probabilities remain in the interval $%
\left[ 0,1\right] $ for all values of k$\in (0,\infty )$.

It is interesting to get the values of the detection probabilities for small
k. Eqs.$\left( \ref{13}\right) $ and $\left( \ref{15}\right) $ give 
\begin{eqnarray*}
P_{A} &=&P_{B}=k\frac{\left| \gamma \right| ^{2}}{2}-k^{2}\frac{\left|
\gamma \right| ^{4}}{8}+\frac{k^{2}}{4\alpha }\left( 1+\left| \gamma \right|
^{4}\right) +O\left( k^{3}\right) , \\
P_{AB} &=&k^{2}\frac{\left| \gamma \right| ^{4}}{4}+O\left( k^{3}\right)
\simeq P_{A}P_{B}
\end{eqnarray*}
The result is that the single detection probability is of order O(k) while
the coincidence detection is of order O(k$^{2}$). This is plausible because
k may be interpreted as a detection efficiency parameter. The latter
equality means that the single detections are uncorrelated at low
efficiencies.

For high values of the parameter k both single and coincidence probabilities
approach unity. Therefore the ratio between coincidence and single
probabilties is unity, which agrees with the quantum prediction eqs.$\left( 
\ref{ra}\right) $ and $\left( \ref{rab}\right) .$

The main result of the calculation in this section is that our model
predicts an increase of the ratio between coincidence and single detection
rates when the parameter k increases. For small k the coincidence detection
probability is about the product of the single probabilities, which implies
that single detections are uncorrelated. For large k the coincidence
probability approaches the single one, which implies a positive correlation
of the singles. Of course for very large k both single and coincidence
probilities become unity, there is detection in every time window.

\textbf{Summary and conclusions of section 4}

I have shown that the Wigner (or Weyl-Wigner, WW) representation of the
quantum electromagnetic field interacting with macroscopic objects allows
suggests local models of experiments, in particular those measuring the
correlation between ``entangled photon pairs'', provided that we supplement
the WW formalism with ``measurement postulates'' consisiting of rules for
the probability of detection as a function of the arriving intensities at
the photodetectors.

Crucial for the agreement with standard quantum predictions is to take due
account of the vacuum radiation (ZPF) arriving at the detectors in addition
to the signal radiation coming from the source in the correlation
experiments. The effect of the ZPF was the subject of subsection 4.2.

For the sake of clarity in subsection 4.3 I have contrived a model (not
realistic) that closely parallels the quantum treatment of correlation
experiments with entangled photon pairs made in section 3. A by-product of
the subsection is an intuitive understanding of entanglement as a
correlation between signals and vacuum ZPF field.

In addition to being local the models will be realistic if the detection
probabilities are physical, in particular they agree with the constraints of
probability theory. Also we should take into account that light beams
consists of many, not just a few, modes of tje radiation field (plane
waves). Both these requirements are included in the realistic local model of
subsection 4.4.

The results of this section for a simple experiment with entangled photon
pairs \textit{suggest }that also realistic local models might be possible
for experiments testing Bell inequalities \cite{Shalm}, \cite{Giustina}.
However there are big difficulties in this case which will be studied in the
next section 5.

\section{On the refutation of local realism by experiments}

The question whether local realistic models are possible for experiments
violating a Bell inequality is most relevant in present day research in
physics. The violation of the inequalities seems to contradict relativistic
causality, thus showing a dramatic conflict between the two fundamental
theories of physics: relativity and quantum mechanics. There are also
technological implications because the said violation supports the current
opinion that nature does not obbey the classical laws of physics. Therefore
there may be technologies which are not possible if nature is governed by
classical laws, but possible otherwise. For instance quantum computers.

In this section we study the question firstly revisiting the work of Bell in
subsection 5.1. Then I will present a local, but not realistic, model of the
experiments violating a Bell inequalities, in subsection 5.2. Finally I
study in subsection 5.3 whether the model might be extended to the point of
violating a Bell inequality. If this was the case there would be still
loopholes in the refutation of local realism by experiments, against the
current view.

\subsection{Local realism and the Bell inequalities}

Local realism has been defined by H. Wiseman as the assumption that: ``The
world is made up of real stuff, existing in space and changing only through
local interactions. This local-realism hypothesis is about the most
intuitive scientific postulate imaginable'' \cite{Wiseman}. Against that
intuition local realism has been questioned since 1964, when Bell proposed
his celebrated inequalities \cite{Bell}. From that time on the scientific
community has accepted the dilemma put by Bell: either quantum mechanics is
incorrect or local realism is untenable. The decission for one of the terms
of the dilemma could be solved by experiments: any (loophole-free) violation
of a Bell inequality should refute local realism. The resulsts of the
experiments \cite{Shalm}, \cite{Giustina} have lead to the belief that local
realism has been actually refuted \cite{Aspect}. The aim of this paper is to
further study whether the belief is correct.

As is well known Bell \cite{Bell64} defined ``local hidden variables
models'', later named ``local realistic'', those models of an experiment
where the results of all correlation measurements amongst two parties (Alice
and Bob) may be interpreted according to the formulas

\begin{eqnarray}
\left\langle \theta \right\rangle &=&\int \rho \left( \lambda \right)
d\lambda M_{A}\left( \lambda ,\theta \right) ,\left\langle \phi
\right\rangle =\int \rho \left( \lambda \right) d\lambda M_{B}\left( \lambda
,\phi \right) ,  \nonumber \\
\left\langle \theta \phi \right\rangle &=&\int \rho \left( \lambda \right)
d\lambda M_{A}\left( \lambda ,\theta \right) M_{B}\left( \lambda ,\phi
\right) ,  \label{bell}
\end{eqnarray}
with standard notation. The quantities $\left\langle \theta \right\rangle
,\left\langle \phi \right\rangle $ and $\left\langle \theta \phi
\right\rangle $ are the expectation values for the results of measuring the
observables $\theta ,\phi $ and their product $\theta \phi $, respectively.
In this paper I will consider that the observables correspond to detection,
or not, of some signals (currently named ``photons'') by either Alice or
Bob, attaching the values 1 or 0 to those two possibilities. In this case $%
\left\langle \theta \right\rangle ,\left\langle \phi \right\rangle $ refer
to the single, and $\left\langle \theta \phi \right\rangle $ to the
coincidence, detection probabilities in a trial of the experiment, all runs
made in identical conditions. The reported rates are the products of these
probabilities times the rate of trials. The following mathematical
conditions are assumed 
\begin{equation}
\rho \left( \lambda \right) \geq 0,\int \rho \left( \lambda \right) d\lambda
=1,M_{A}\left( \lambda ,\theta \right) \in \left[ 0,1\right] ,M_{B}\left(
\lambda ,\phi \right) \in \left[ 0,1\right] .  \label{bell1}
\end{equation}

A constraint of locality is included, namely $M_{A}\left( \lambda ,\theta
\right) $ should be independent of the choice of the observable $\phi $, $%
M_{B}\left( \lambda ,\phi \right) $ independent of $\theta $ and $\rho
\left( \lambda \right) $ independent of both $\theta $ and $\phi $ . From
these conditions it is possible to derive empirically testable (Bell)
inequalities, e.g. the one derived by Clauser and Horne \cite{CH}. The tests
are most relevant if the measurements performed by Alice and Bob are
spacially separated in the sense of relativity theory, because in this case
the violation of a Bell inequality would refute relativistic causality.

For experiments measuring polarization correlation of photon pairs the CH
inequality \cite{CH} may be written

\begin{equation}
C\equiv \left\langle \theta _{1}\right\rangle +\left\langle \phi
_{1}\right\rangle +\left\langle \theta _{2}\phi _{2}\right\rangle
-\left\langle \theta _{1}\phi _{1}\right\rangle -\left\langle \theta
_{1}\phi _{2}\right\rangle -\left\langle \theta _{2}\phi _{1}\right\rangle
\geq 0,  \label{CH}
\end{equation}
where $\theta _{j}$ stands for the observable ``detection event by the Alice
detector placed in front of a polarizer at angle $\theta _{j}$''. Similarly $%
\phi _{k}$ for Bob detector. The proof is simple and well know but in the
following I exhibit it in order to show the strength of the inequality.
Indeed if we have for real numbers fulfilling 
\[
\theta _{1},\phi _{1},\theta _{2},\phi _{2}\in \left[ 0,1\right] , 
\]
a trivial mathematical result is that the following inequality holds true 
\[
\theta _{1}+\phi _{1}+\theta _{2}\phi _{2}\geq \theta _{1}\phi _{1}+\theta
_{1}\phi _{2}+\theta _{2}\phi _{1}. 
\]
Multiplication times the positive definite quantity $\rho \left( \lambda
\right) $ and integration with respect to $\lambda $ gives eq.$\left( \ref
{CH}\right) .$

Ten years after his original formulation, i.e. eqs.$\left( \ref{bell}\right)
,$ Bell went further \cite{Bell75} proposing another formulation of local
realism resting on assumptions more general, although less formalized$.$ His
proposal was that all correlations between the results of distant
measurements (by Alice and Bob) derive from events in their common past
light cone. Hence he derived a ``locality inequality'' practically
equivalent to the Bell inequalities already known. After that there is a
widely held belief that any true experimental violation of a Bell inequality
contradicts locality, now understood as relativistic causality.

\subsection{Local model for correlations of photons entangled in polarization
}

In the following I work a local model for an experiment measuring the
correlation of photons entangled in polarization produced via SPDC. The
loophole-free experiments violating a Bell inequality are of that class \cite
{Shalm}, \cite{Giustina}. Typically the experiment consists of a source
producing light beams that are sent to two detection systems. The analysis
is similar to the one made for the signal-idler correlation in section 4.3
although more lenghty. Here I will\textit{\ }use the (non-physical)
detection assumption eq.$\left( \ref{2I}\right) ,$ that is detection
probability proportional to intensity. However when a physical detection is
used, i.e. fulfilling eqs.$\left( \ref{51}\right) $ and $\left( \ref{52}%
\right) ,$the ratio between coincidence and single detection is smaller. The
reason is that the physical detection rules diminish the detection
probability for high intensities due to the constraint eq.$\left( \ref{52}%
\right) $. Hence the coincidence probabilities decrease (twice times) more
that single probabilities, similar to the example exhibited in section 4.4.
The consequence that inequalities like eq.$\left( \ref{CH}\right) $ are
never violated.

Anyway the question is whether local realistic models exist for the
empirical tests showing the violation of a Bell inequality. This question
will be discussed again in the next section 5.3, here I will just show that
using the non-physical detection rule eq.$\left( \ref{2I}\right) $ there is
a model that reproduces the ideal quantum prediction and does violate a Bell
inequality.

In the case of photons entangled in polarization we need 4 scalar modes
combining two different wavevectors and two perpendicular linear
polarizations each. I omit the HS and WW analyses of the experiment, see 
\cite{FOOP}, and go directly to a local model as described for the more
simple experiment in subsection 4.3. But I point out that \textit{the model
is local but not realistic} because the non-physical detection rule eq.$%
\left( \ref{2I}\right) $ is used. Therefore it does not disprove the claimed
refutation of local realism\textit{. }The interest of the model is to
provide clues in the search of true local realistic models, as commented on
the following subsection 5.3.

I will represent every two scalar modes with the same wavevector as a single
vector mode with a 2D vector amplitude, represented in boldface. Then by
analogy with eq.$\left( \ref{pairs}\right) $ I will write

\begin{equation}
\mathbf{a}_{s}\mathbf{\rightarrow b}_{s}\mathbf{\equiv a}_{s}\mathbf{+}%
\gamma \mathbf{a}_{i}^{*}\mathbf{,a}_{i}\mathbf{\rightarrow b}_{i}\mathbf{%
\equiv a}_{i}\mathbf{+}\gamma \mathbf{a}_{s}^{*},  \label{pairs2}
\end{equation}
where $\mathbf{a}_{s}$ and $\mathbf{a}_{i}$ is the relevant radiation that
enter the crystal and $\mathbf{b}_{s},\mathbf{b}_{i}$ the outgoing radiation.

The beams $\mathbf{b}_{s},\mathbf{b}_{i}$ are combined in a polarizing beam
splitter so that the vector field amplitudes sent to Alice, $\mathbf{E}_{A}$%
, and Bob, $\mathbf{E}_{B},$ respectively, may be represented as follows

\begin{equation}
\mathbf{E}_{A}=\mathbf{b}_{sx}+i\mathbf{b}_{iy},\mathbf{E}_{B}=\mathbf{b}%
_{sy}+i\mathbf{b}_{ix}.  \label{pairs3}
\end{equation}
For the sake on clarity I shall specify that $\mathbf{E}_{A}$ is a mode
(plane wave) with wavevector $\mathbf{k}_{A}$ having the direction of the
straight line going from the crystal to Alice. It has a component, $\mathbf{b%
}_{sx}$, say horizontal, and another component, $i\mathbf{b}_{iy}$,
vertical, and similar for Bob. An alternative notation would be 
\[
\mathbf{E}_{A}=(b_{sx},ib_{iy}),\mathbf{E}_{B}=(b_{sy},ib_{ix}). 
\]
Alice has a polarization analyzer at an angle $\theta $ with the horizontal
and a detector in front of it, and similarly Bob with the analyzer at an
angle $\phi $. Thus the (scalar) amplitudes arriving at the detectors are,
respectively, 
\begin{equation}
E_{A}=b_{sx}\cos \theta +ib_{iy}\sin \theta ,E_{B}=ib_{sx}\cos \phi
+b_{iy}\sin \phi .  \label{pairs4}
\end{equation}
The field amplitude $E_{A}$ ($E_{B})$ arriving at Alice (Bob) detector is
the $\theta $ $\left( \phi \right) $ component of the vector field amplitude 
$\mathbf{E}_{A}$ $(\mathbf{E}_{B}).$

In order to proceed with the analysis of the experiment it is convenient to
write eqs.$\left( \ref{pairs4}\right) $ in terms of the initial amplitudes,
in analogy with eqs.$\left( \ref{pairs2}\right) ,$ that is 
\begin{eqnarray}
E_{A} &=&(a_{sx}+\gamma a_{ix}^{*})\cos \theta +i(a_{iy}+\gamma
a_{sy}^{*})\sin \theta  \nonumber \\
&=&a_{sx}\cos \theta +ia_{iy}\sin \theta +\gamma \left( a_{ix}^{*}\cos
\theta +ia_{sy}^{*}\sin \theta \right) \equiv c_{1}+\gamma c_{2}^{*} 
\nonumber \\
E_{B} &=&ia_{ix}\cos \phi +a_{sy}\sin \phi +\gamma \left( ia_{sx}^{*}\cos
\phi +a_{iy}^{*}\sin \phi \right) \equiv d_{2}+\gamma d_{1}^{*}.  \label{5M}
\end{eqnarray}
Hence the intensities arriving to Alice and Bob detectors will be,
respectively, 
\begin{eqnarray}
I_{A} &=&\left| c_{1}\right| ^{2}+\left| \gamma \right| ^{2}\left|
c_{2}\right| ^{2}+Y_{A}=\left| \gamma \right| ^{2}\left| c_{2}\right|
^{2}+Z_{A},  \nonumber \\
I_{B} &=&\left| d_{2}\right| ^{2}+\left| \gamma \right| ^{2}\left|
d_{1}\right| ^{2}+Y_{B}=\left| \gamma \right| ^{2}\left| d_{1}\right|
^{2}+Z_{B},  \label{M5}
\end{eqnarray}
neglecting terms like $a_{sx}^{2}$ and $a_{sx}a_{iy}$ rapidly fluctuating,
so that there are no terms linear in $\gamma .$ The latter term is the
contribution of the ZPF, in analogy with eqs.$\left( \ref{M2}\right) .$

Eqs.$\left( \ref{M5}\right) $ are similar to eqs.$\left( \ref{M2}\right) ,$
the difference being that now $c_{1}$ and $d_{2}$ are linear combinations of
the (``signal'' and ``idler'') components of vacuum vector amplitudes that
entered the crystal and emerged from it without change. In contrast the
amplitudes involved in eqs.$\left( \ref{M2}\right) ,$ $a_{s},a_{i},$ were
simply vacuum scalar amplitudes that crossed the crystal. Similarly $%
c_{1}^{*}$, $d_{2}^{*}$ are linear combinations of components of the
amplitudes produced in the crystal by the action of the laser while $%
a_{s}^{*},a_{i}^{*}$ were simply the amplitudes produced in the crystal by
the action of the laser. The quantities Y and Z are effective intensities
resulting from the ZPF arriving at the Alice and Bob detectors, quite
similar to eqs.$\left( \ref{M2}\right) .$

Our assumption for the detection probabilities will be 
\begin{equation}
P_{A}=\left\langle I_{A}\right\rangle ,P_{B}=\left\langle I_{B}\right\rangle
,P_{AB}=\left\langle I_{A}I_{B}\right\rangle .  \label{M8}
\end{equation}
The calculation is parallel to that in section 4.3 for the correlation
between signal and idler. The single detection probability by Alice will be,
taking eq.$\left( \ref{M5}\right) $ into account, 
\begin{eqnarray}
P_{A} &=&\left\langle \left| \gamma \right| ^{2}\left| c_{2}\right|
^{2}+Z_{A}\right\rangle =\left| \gamma \right| ^{2}\left\langle \left|
c_{2}\right| ^{2}\right\rangle =\left| \gamma \right| ^{2}\left\langle
\left| a_{ix}^{*}\cos \theta +ia_{sx}^{*}\sin \theta \right|
^{2}\right\rangle  \nonumber \\
&=&\left| \gamma \right| ^{2}\left[ \left\langle \left| a_{ix}^{*}\right|
^{2}\right\rangle \cos ^{2}\theta +\left\langle \left| ia_{sx}^{*}\right|
^{2}\right\rangle \sin ^{2}\theta \right] =\frac{1}{2}\left| \gamma \right|
^{2},  \label{M6} \\
P_{B} &=&\left| \gamma \right| ^{2}\left\langle \left| d_{1}\right|
^{2}\right\rangle =\left| \gamma \right| ^{2}\left\langle \left|
ia_{ix}^{*}\cos \phi +a_{sx}^{*}\sin \phi \right| ^{2}\right\rangle =\frac{1%
}{2}\left| \gamma \right| ^{2},  \nonumber
\end{eqnarray}
where $\left\langle Z_{A}\right\rangle =\left\langle Z_{B}\right\rangle =0$
and we have taken into account the following averages, weighted by eq.$%
\left( \ref{1}\right) ,$%
\begin{eqnarray}
\left\langle \left| a_{sx}\right| ^{2}\right\rangle &=&\left\langle \left|
a_{ix}\right| ^{2}\right\rangle =\left\langle \left| a_{sx}\right|
^{4}\right\rangle =\left\langle \left| a_{ix}\right| ^{4}\right\rangle =%
\frac{1}{2},  \nonumber \\
\left\langle a_{sx}^{2}\right\rangle &=&\left\langle a_{ix}^{2}\right\rangle
=\left\langle a_{sx}a_{ix}\right\rangle =\left\langle
a_{sx}^{*}a_{ix}^{*}\right\rangle =0.  \label{Ave}
\end{eqnarray}
The coincidence detection probability should be obtained from the former
expressions of eqs.$\left( \ref{M5}\right) $ in analogy with eqs.$\left( \ref
{AB}\right) $ and $\left( \ref{Q1}\right) .$ We get 
\begin{eqnarray}
P_{AB} &=&\left\langle I_{A}I_{B}\right\rangle =\left\langle \left( \left|
c_{1}\right| ^{2}+\left| \gamma \right| ^{2}\left| c_{2}\right|
^{2}+Y_{A}\right) \left( \left| d_{2}\right| ^{2}+\left| \gamma \right|
^{2}\left| d_{1}\right| ^{2}+Y_{B}\right) \right\rangle  \nonumber \\
&=&\left\langle \left| c_{1}\right| ^{2}\right\rangle \left\langle \left|
d_{2}\right| ^{2}\right\rangle +\left| \gamma \right| ^{4}\left\langle
\left| c_{2}\right| ^{2}\right\rangle \left\langle \left| d_{1}\right|
^{2}\right\rangle +\left| \gamma \right| ^{2}\left( \left\langle \left|
c_{2}\right| ^{2}\left| d_{2}\right| ^{2}\right\rangle +\left\langle \left|
c_{1}\right| ^{2}\left| d_{1}\right| ^{2}\right\rangle \right)  \nonumber \\
&&-\left\langle \left| c_{1}\right| ^{2}\right\rangle \left\langle \left|
d_{2}\right| ^{2}+\left| \gamma \right| ^{2}\left| d_{1}\right|
^{2}\right\rangle -\left\langle \left| c_{1}\right| ^{2}+\left| \gamma
\right| ^{2}\left| c_{2}\right| ^{2}\right\rangle \left\langle \left|
d_{2}\right| ^{2}\right\rangle +\left\langle \left| c_{1}\right|
^{2}\right\rangle \left\langle \left| d_{2}\right| ^{2}\right\rangle 
\nonumber \\
&=&\left| \gamma \right| ^{4}\left\langle \left| c_{2}\right|
^{2}\right\rangle \left\langle \left| d_{1}\right| ^{2}\right\rangle +\left|
\gamma \right| ^{2}\left( \left\langle \left| c_{2}\right| ^{2}\left|
d_{2}\right| ^{2}\right\rangle -\left\langle \left| c_{2}\right|
^{2}\right\rangle \left\langle \left| d_{2}\right| ^{2}\right\rangle \right)
\nonumber \\
&&+\left| \gamma \right| ^{2}\left( \left\langle \left\langle \left|
c_{1}\right| ^{2}\left| d_{1}\right| ^{2}\right\rangle \right\rangle
-\left\langle \left| c_{1}\right| ^{2}\right\rangle \left\langle \left|
d_{1}\right| ^{2}\right\rangle \right)  \nonumber \\
&=&\frac{1}{4}\left| \gamma \right| ^{4}+\left| \gamma \right| ^{2}\left[
\left( \left\langle \left| c_{2}\right| ^{2}\left| d_{2}\right|
^{2}\right\rangle -\frac{1}{4}\right) +\left( \left\langle \left|
c_{1}\right| ^{2}\left| d_{1}\right| ^{2}\right\rangle -\frac{1}{4}\right)
\right] .  \label{7M}
\end{eqnarray}
The term $\left\langle \left| c_{2}\right| ^{2}\left| d_{2}\right|
^{2}\right\rangle $ may be got writing the fields $c_{2}$ and $d_{2}$ in
terms of the amplitudes $a$ and then taking eqs.$\left( \ref{Ave}\right) $
into account in order to obtain the averages. The algebra is simplified
using the well known rule for the average of the product of four Gaussian
variables, that is 
\[
\left\langle ABCD\right\rangle =\left\langle AB\right\rangle \left\langle
CD\right\rangle +\left\langle AC\right\rangle \left\langle BD\right\rangle
+\left\langle AD\right\rangle \left\langle BC\right\rangle . 
\]
We get 
\begin{eqnarray*}
\left\langle \left| c_{2}\right| ^{2}\left| d_{2}\right| ^{2}\right\rangle
&=&\left\langle c_{2}c_{2}^{*}\right\rangle \left\langle
d_{2}d_{2}^{*}\right\rangle +\left\langle c_{2}d_{2}\right\rangle
\left\langle c_{2}^{*}d_{2}^{*}\right\rangle +\left\langle
c_{2}d_{2}^{*}\right\rangle \left\langle c_{2}^{*}d_{2}\right\rangle \\
&=&\frac{1}{4}+0+\left| \frac{1}{2}\cos \theta \cos \phi +\frac{1}{2}\sin
\theta \sin \phi \right| ^{2} \\
&=&\frac{1}{4}+\frac{1}{4}\cos ^{2}\left( \theta -\phi \right) .
\end{eqnarray*}
A similar result is got for $\left\langle \left| c_{1}\right| ^{2}\left|
d_{1}\right| ^{2}\right\rangle $ whence the coincidence detection
probability becomes 
\begin{eqnarray}
P_{AB} &=&\frac{1}{2}\left| \gamma \right| ^{2}\cos ^{2}\left( \theta -\phi
\right) +O(\left| \gamma \right| ^{4})  \label{M7} \\
&=&\frac{1}{4}\left| \gamma \right| ^{2}\left[ 1+\cos \left( 2\theta -2\phi
\right) \right] +O(\left| \gamma \right| ^{4}).  \nonumber
\end{eqnarray}

The predictions of the local model, eqs.$\left( \ref{M6}\right) $ and $%
\left( \ref{M7}\right) ,$ violate Bell inequalities for appropriate choices
of the angles. In fact it we insert the model predictions eqs.$\left( \ref
{M6}\right) $ and $\left( \ref{M7}\right) $ in eq.$\left( \ref{CH}\right) $
we get 
\begin{eqnarray}
C &=&\eta \left| \gamma \right| ^{2}+\eta ^{2}(-\frac{1}{2}\left| \gamma
\right| ^{2}+\frac{1}{4}\left| \gamma \right| ^{2}D-\frac{1}{2}\left| \gamma
\right| ^{4})  \nonumber \\
D &\equiv &\left[ \cos \left( 2\theta _{2}-2\phi _{2}\right) -\cos \left(
2\theta _{1}-2\phi _{1}\right) -\cos \left( 2\theta _{1}-2\phi _{2}\right)
-\cos \left( 2\theta _{2}-2\phi _{1}\right) \right] ,  \label{nobell}
\end{eqnarray}
where I have included the efficiency $\eta $ of detectors, assumed the same
for both detectors. We get C\TEXTsymbol{<}0 violating the inequality $\left( 
\ref{CH}\right) $ for some choices of the angles and large enough efficiency 
$\eta $. For instance if 
\[
\theta _{1}=\frac{\pi }{2},\theta _{2}=0,\phi _{1}=\frac{\pi }{4},\phi _{2}=%
\frac{3\pi }{4}, 
\]
we get 
\[
C=\eta \left| \gamma \right| ^{2}-\eta ^{2}\left( \frac{\sqrt{2}+1}{2}\left|
\gamma \right| ^{2}+O(\left| \gamma \right| ^{4})\right) , 
\]
where $C<0$ for 
\[
\eta >\left( \frac{\sqrt{2}+1}{2}+\frac{1}{2}\left| \gamma \right|
^{2}\right) ^{-1}\simeq 0.828\text{ if }\left| \gamma \right| ^{2}<<1. 
\]

I stress again that the results eqs.$\left( \ref{M6}\right) $ and $\left( 
\ref{M7}\right) $ have not been obtained with the detection assumption eq.$%
\left( \ref{51}\right) $ and therefore it may not fulfil the probabilty
constraint eq.$\left( \ref{52}\right) .$ Therefore the model worked in this
section \textit{is not an actual realistic local model }of the experiments
violating a Bell inequality. the non-physical detection rule eq.$\left( \ref
{2I}\right) .$ Consequently \textit{it does not disprove the claimed
refutation of local realism.}

\subsection{Do optical local models exist violating a Bell inequality?}

Two methods exist in order to try disproving the current opinion that local
realism has been refuted by performed experiments \cite{Shalm}, \cite
{Giustina} : 1) to find a new loophole showing the these experiments are not
really loophole-free, 2) to propose a counter example to the claim that Bell
inequalities should be fulfilled by all local realistic models, that is to
construct a model violating a Bell inequality. In this subsection I will be
concerned with the second possibility.

In order to answer the question there are arguments \textit{for and against}
a positive answer. The main argument \textit{against} the possibility is
that Bell\'{}s derivation of the inequalities is quie general. An argument 
\textit{for} the said possibility is the existence of models, like that of
the previous section 5.2, which are suggested by the Wigner representation
(WW) and agree with quantum predictions for low effective detection
efficiencies, although not for the high efficiencies required for the
violation of a Bell inequality. Here effective efficiency means roughly the
ratio between typical coincidence and single detection probabilities.

Let us discuss more carefully the arguments \textit{for} local models. The
analysis of correlation experiments within the Hilbert space formalism (HS),
as in subsection 4.2.1, does not provide any clue for the construction of LR
models. In particular eq.$\left( \ref{qpairs}\right) $ contains an operator $%
\hat{a}_{s}$ associated to a relatively large radiation amplitude and
another operator $\gamma \hat{a}_{i}^{\dagger }$ associated to a weaker
amplitude (weaker because $\left| \gamma \right| <<1).$ However the former
amplitude does not contribute to detection, it appears just as an auxiliary
element in the formalism, but the latter does contribute, which is essencial
for the agreement with experiments. The different role of a creation and a
destruction operator is not easy to understand intuitively, with the result
that the standard HS does not provide any clue for the construction of LR
models. In sharp contrast the Wigner representation (WW) suggests an
intuitive picture. That is the intensity \textit{I}$_{s}=\left| a_{s}\right|
^{2}$ is a part of the ZPF, whose effect on detectors is precisely balanced
by the rest of the ZPF. Thus WW does suggest how to construct LR, that is
taking the vacuum field ZPF into account. This fact has been exploited in
the models quoted in the Introduction section \cite{Frontiers}, \cite{FOOP}, 
\cite{book} or those worked in subsections 4.3 and 5.1 above. However these
models are not realistic because an unphysical detection rule is used. A
true LR model has been presented in section 4.4 for a simple experiment able
to reproduce the quantum predictions, but it is not obvious that a similar
model may be devised for the experiments testing a Bell inequality. Hence I
cannot find compulsory arguments for a positive answer to the question posed
in the title of this subsection.

The main argument \textit{against} the possibility of LR models is the
generality of Bell definition of local realism, a definition which leads
strightforward to the inequalities. Indeed it is the case that Bell\'{}s
definition of local realism, eq.$\left( \ref{bell1}\right) ,$ is general
enough to include the possible action of the ZPF mentioned in the previous
paragraph, because the distribution function $\rho \left( \lambda \right) $
may include \textit{all }random variables of the radiation arriving at a
detector, that is also the vacuum ZPF. The conclusion is that the inclusion
of the ZPF does not make easier the construction of local realistic models.

Up to here it seems that the arguments against the possibility of LR models
are uncontroversial. Nevertheless there is a weak point in the definition eq.%
$\left( \ref{bell1}\right) $ of local realism. It is not in the function $%
\rho \left( \lambda \right) ,$ but in the response functions $M_{A}(\lambda
,\theta )$ and $M_{B}\left( \lambda ,\phi \right) .$ They have the flavor of
the quantum measurement theory, whose mere existence was disputed in section
2.2. For instance $M_{A}(\lambda ,\theta )$ means that ``if we measure (e.g.
we detect) a signal arriving at Alice the response is \textit{yes} for some
values of $\lambda $ and \textit{no} for other values''. However this is an
extreme idealization of the detection process. In fact in a realistic
approach the detection is a complex process involving the interaction of the
incoming radiation with some resonant material system inside the detector.
The process is not instantaneous but it requires a time interval $T_{\det }$
large in comparison with the period of the incoming radiation, that is $%
T_{\det }>>\omega ^{-1}.$ In fact $T_{\det }$ may depend on the distribution
of frequencies in the incoming beam, and we know that a narrow beam would
consist of the superposition of many waves with different wavevectors. All
this complexity cannot be summarized in a single parameter of detection
efficiency. In the following I give a simple illustrative example.

Assuming that we can just know whether in a given run of the experiment
there is at least one detection or none, say by Alice. That is we cannot
distinguish one from several detection events within a given time window.
Let T\TEXTsymbol{>}$T_{\det }$ be the duration of the time window chosen for
a run of the experiment, and let \textit{p} be the probability of single
detection by either Alice or Bob in the first half window with duration T/2.
An similar for the second half. Let q\TEXTsymbol{<}p be the probability of
coincidence either in this first half or in the second half, and let r the
probability of a coincidence due to detection by Alice in the first half and
by Bob in the second or viceversa\textbf{\ }excluding coincidence either in
the first or the second half or both. Then the single probability of at
least one count by Alice in either the first or the second half window will
be 
\[
2p-p^{2}
\]
and similar for Bob. The coincidence detection probability will be 
\[
2q-q^{2}+r.
\]
Then the ratio coincidence/single in the total time window is greater than
the ratio in one of the half windows. That is the difference 
\begin{equation}
\frac{2q-q^{2}+r}{2p-p^{2}}-\frac{q}{p}=\frac{q}{p}\times \frac{2-q}{2-p}+%
\frac{r}{2p-p^{2}}>0,  \label{5a}
\end{equation}
because it is the sum of two non-negative quantities. Hence we conclude that
if the time window for a run of the experiment is large and we do not
distinguish between one or two detection events within that window, the
ratio between coincidences and singles will be greater than for shorter time
windows. That is the sign of the quantity C in eq.$\left( \ref{CH}\right) $
may depend on the length of the duration of the time window. And
consequently the violation, or not, of the Bell inequality eq.$\left( \ref
{CH}\right) .$

This has been a simple illustrative example of the ambiguity of the response
functions $M_{A}(\lambda ,\theta )$ and $M_{B}\left( \lambda ,\phi \right) $
in eqs.$\left( \ref{bell1}\right) .$ More generally the point is that the
response in eqs.$\left( \ref{bell1}\right) $ \textit{should not be functions
but functionals}. We should write for the responses $M_{A}\left[ \lambda
\left( t\right) ,\theta \right] $ and $M_{B}\left[ \lambda \left( t\right)
,\phi \right] $ for t$\in \left[ t^{\prime },t^{\prime }+T\right] ,$ t'
being the starting time of the window and t'+T the final time.

\textbf{Conclusions of section 5.}

Bell\'{}s definition of local realism eqs.$\left( \ref{bell1}\right) $ is
not general enough in the case of optical experiments, involving
electromagnetic signals. Then a local realistic model violating a Bell
inequality might be possible. That is a model like that of subsection 4.3
but with modified detection assumptions. Searching for such a model is
worthwhile, taking into account the far reaching consequences of a conflict
between quantum mechanics and relativity theory.

\end{document}